\documentclass[a4paper,11pt]{article}

\usepackage{amsmath}
\usepackage{graphicx}
\usepackage{amsfonts}
\usepackage{amssymb}
\usepackage{amsmath}
\usepackage{amsthm}
\usepackage{dsfont}
\usepackage{mathtools}
\usepackage{empheq}
\usepackage{bm}
\usepackage{ulem}  
\usepackage[svgnames]{xcolor}
\usepackage{stmaryrd}
\usepackage{tikz}
\usepackage[nomessages]{fp}
\usepackage{color}
\usepackage{url}
\usepackage{empheq}

\usepackage{aliascnt}
\usepackage{appendix}
\usepackage[unicode]{hyperref}
 
\newcommand{\mybullet}{{\hbox{\scalebox{.4}{$\bullet$}}}}
\newcommand{\ind}[1]{\mathbf{1}_{#1}}
\theoremstyle{plain}
\newtheorem{defi}{Definition}
\newtheorem{lem}[defi]{Lemma}
\newtheorem{coro}[defi]{Corollary}
\newtheorem{prop}[defi]{Proposition}
\newtheorem{thm}[defi]{Theorem}
\theoremstyle{definition}

\theoremstyle{remark}

\definecolor{orange-lpens}{RGB}{230,62,37}
\definecolor{bleu-lpens}{RGB}{41,50,117}

\newcommand{\bc}{\begin{center}}
\newcommand{\ec}{\end{center}}
\newcommand{\beq}{\begin{equation}}
\newcommand{\eeq}{\end{equation}}
\newcommand{\beqa}{\begin{eqnarray}}
\newcommand{\eeqa}{\end{eqnarray}}
\newcommand{\beqs}{\begin{eqnarray*}}
\newcommand{\eeqs}{\end{eqnarray*}}

\newcommand{\bi}{\begin{itemize}}
\newcommand{\ei}{\end{itemize}}

\newcommand{\bra}{\langle}
\newcommand{\ket}{\rangle}
\newcommand{\Tr}{{\rm Tr}}

\def\ket#1{|#1\rangle}
\def\bra#1{\langle#1|}

\newcommand{\oo}{\text{\o}}
\def\cadremath#1{\vbox{\hrule\hbox{\vrule\kern8pt\vbox{\kern8pt
			\hbox{ {$\displaystyle #1 $ } }\kern8pt} 
			\kern8pt\vrule}\hrule}}
\title{\bf Bernoulli variables, classical exclusion processes and free probability}
\author{Michel Bauer${}^{\blacklozenge}$\footnote{michel.bauer@ipht.fr}, Denis Bernard${}^{\clubsuit}$\footnote{denis.bernard@ens.fr}, Philippe Biane${}^{\spadesuit}$\footnote{biane@univ-mlv.fr} and Ludwig Hruza${}^{\clubsuit}$\footnote{ludwig.hruza@ens.fr}}
\date{\today}

\begin{document}

\maketitle
\pagestyle{empty}


{\centering{\small{$^\blacklozenge$ Universit\'e Paris-Saclay, CNRS, CEA, Institut de Physique Th\'eorique, 91191 Gif-sur-Yvette, France; PSL Research University, CNRS, \'Ecole normale sup\'erieure, D\'epartement de math\'ematiques et applications, 75005 Paris, France
 }}}

{\centering{\small{$^\clubsuit$ Laboratoire de Physique de l'\'Ecole Normale Sup\'erieure, CNRS, ENS \& Universit\'e PSL, Sorbonne Universit\'e, Universit\'e Paris Cit\'e, 75005 Paris, France.}}}

{\centering{\small{$^\spadesuit$ Laboratoire d'Informatique Gaspard Monge, CNRS, Universit\'e Gustave Eiffel,
77454 Champs-Sur-Marne,
France  }}}
\vskip 0.5 truecm

\pagestyle{plain}

\begin{abstract}
 We present a new description of the known large deviation function of the classical symmetric simple exclusion process by exploiting its connection with the quantum symmetric simple exclusion processes and using tools from free probability. This may seem paradoxal as free probability usually deals with non commutative probability while the simple exclusion process belongs to the realm of classical probability. On the way, we give a new formula for the free energy -- alias the logarithm of the Laplace transform of the probability distribution -- of correlated Bernoulli variables in terms of the set of their cumulants with non-coinciding indices. This latter result is obtained either by developing a combinatorial approach for cumulants of products of random variables or by borrowing techniques from Feynman graphs.
\end{abstract}

\medskip

\centerline{\it In memory of Krzysztof GAWEDZKI (1947-2022)\footnote{\url{https://en.wikipedia.org/wiki/Krzysztof_Gawedzki}}.}
\noindent {\it We imagine that Krzysztof would have appreciated this manuscript, which intertwines problems in physics and mathematics. We hope that it fits with Krzysztof's rigorous, precise exploration of mathematical physics.}

\vskip 0.5 truecm

{\setlength{\parskip}{0pt plus 1pt} \tableofcontents}

\medskip
\centerline{-----------------------------}

\section{Introduction and Summary}

The symmetric simple exclusion process (SSEP) \cite{Derrida_Review,Mallick_Review} is an iconic  model of out-of-equilibrium classical statistical physics \cite{Spohn91,Kipnis99}. It describes particles on a line, hopping to the left and right but with the exclusion rule that two particles can never be at the same place. The SSEP  played an important role in the emergence of the so-called macroscopic fluctuation theory~\cite{MFT}, which is a general, phenomenological framework, suited for  dealing with diffusive out-of-equilibrium classical systems. A quantum version of the classical SSEP~\cite{BernardJin19}, named Q-SSEP, has recently been defined. Q-SSEP extends the SSEP in the sense that it keeps track of possible quantum mechanical interferences but in such a way that the classical SSEP is recovered when looking at the average behavior of quantum mechanical observables. It might play a role in a possible quantum extension of the classical macroscopic fluctuation theory~\cite{Bernard21}. 

Interestingly, free probability techniques play an important role in the study of the Q-SSEP, either in constructing its invariant measure \cite{Biane22} or in deciphering its dynamics~\cite{HruzaBernard22}. Since the classical SSEP is embedded in the quantum SSEP, free probability also plays a role in understanding the known characteristics of SSEP, in particular its large deviation rate function. This may sound surprising as free probability has been introduced in the mathematical literature~\cite{Voiculescu1997,Mingo2017,Speicher2019,Biane2003} in order to deal with non commutative probability while SSEP belongs to the realm of classical probability. The purpose of this manuscript is to explain this hidden role. 

On the way, we solve the following problem, apparently simple but for which we did not find an answer in the literature and which reveals nice connections with combinatorial structures. Let $b_i$, $i=1,\cdots,N$, be a collection of $N$ Bernoulli variables, $b_i=0$ or $1$ and let  $K_n(b_{i_1},\cdots,b_{i_n})$ be their cumulants. We call  non-coincident these cumulants when the indices $i_1,\cdots,i_n$ are pairwise distincts (i.e. there are no $ p\not=q$ such that $i_p=i_q$). Since $b_i^2=b_i$ for all $i$, all other cumulants, and hence the joint distribution, are determined from the non-coincident cumulants. Let $Z[h]:=\mathbb{E}\left[e^{\sum_i h_i b_i}\right]$, be the Laplace transform of the joint distribution of the $b_i$'s. To make contact with physics terminology, we shall call it the partition function. It is clearly fully determined by the non-coincident cumulants, since
\beq
Z[h] = \mathbb{E}\Big[\prod_i(1 + b_i\,e_i)\Big],
\eeq
with $e_i:=e^{h_i}-1$. 

The question is then: How to compactly write $W[h]:=\log Z[h]$, the generating function of the cumulants, including coincident indices, in terms of the non-coincident cumulants~?

Of course, the answer to this question is easy when these variables are  independent, since then the generating function factorizes, $Z_\mathrm{free}[h]=\prod_i[1+g_i\, e_i]$, with $g_i:=\mathbb{E}[b_i]$ the mean of $b_i$, and
\beq \label{eq:W-free}
W_\mathrm{free}[h]=\sum_i \log\big[1+g_i e_i\big].
\eeq
It informs on cumulants at coincident points, say at order two $\mathbb{E}[b_i^2]^c=K_2(b_i,b_i)=g_i(1-g_i)$ or three $\mathbb{E}[b_i^3]^c=K_3(b_i,b_i,b_i)=g_i - 3g_i^2 +2 g_i^3$, and similarly at higher orders.
To later make contact with large deviation rate functions, let $S[n]$ be the Legendre transform of $W[h]$, that is: $S_\mathrm{free}[n]=\max_{\{h_i\}}\left[\sum_i h_in_i-W_\mathrm{free}[h]\right]$, then
\beq \label{eq:I-free}
S_\mathrm{free}[n] = \sum_i \left[ n_i\log\big(\frac{n_i}{g_i}\big) + (1-n_i)\log\big(\frac{1-n_i}{1-g_i}\big) \right].
\eeq

A simple formula such as (\ref{eq:W-free}) does not hold in the correlated case. Nevertheless, as explained in Section \ref{sec:Combinatoire}, $W[h]$ admits a  representation as a sum over bipartite graphs whose weights are determined by the non-coincident cumulants. 

\beq
W[h] = \sum_H\frac{\mu (H^\bullet)}{|\text{Aut}\, H|}\sum_{{\mathcal L}\in Lab(H)}w({\mathcal L}) ,
\eeq
where the sum is over all connected bipartite graphs $H$ with an arbitrary number of black but at most $N$ white vertices, and $\mathcal L$ denotes a labelling of the  white vertices by distinct integer indices in $[1,N]$. To such a labelling $\mathcal L$ is associated  a weight $w(\mathcal L)$ described below, see equations (\ref{eq:weight-L},\ref{Weq}).

Things simplify  in the large $N$ scaling limit if we assume that the non-coincident cumulants scale in a specific way at large $N$. Namely, let us assume that 
\beq \label{eq:scaling}
 K_n(b_{i_1},\cdots,b_{i_n}) = \frac{1}{N^{n-1}} \, \psi_n(\frac{i_1}{N},\cdots, \frac{i_n}{N})\, \big(1 + O(\frac{1}{N})\big),
\eeq
for $\psi_n(x_1,\cdots,x_n)$ a collection of continuous functions  and  $h_i=h(\frac{i}{N})$ for $ h(s)$  a continuous function, then only trees contribute to the graph expansion of the cumulant generating functions and $W[h]\sim NF[h]$ at large $N$. In accordance with physics terminology, we shall call $F[h]$ the free energy (per unit of volume). As explained in Sections \ref{sec:Combinatoire} and \ref{sec:Feynman}, the latter can be determined by solving an extremization problem~:
\beq \label{eq:F-bernoulli}
F[h]= \max_{g(\cdot); q(\cdot)} \left[\int_0^1 \!\!\! dx\big[ \log\big(1+g(x)e(x)\big) - q(x)g(x)\big] + F_0[q]\right] ,
\eeq
with $e(x):=e^{h(x)}-1$ and $F_0[q]$ the generating function of non-coincident cumulants,
\beq
F_0[q]:= \sum_{n\geq1}\frac{1}{n!}\int_0^1\! \psi_n(x_1,\cdots,x_n) \prod_{k=1}^n q(x_k)dx_k \,.
\eeq
The extremization problem \eqref{eq:F-bernoulli} has to be solved over all functions $g$ and $q$ on $[0,1]$, without specified boundary conditions. Comparing with the free formula \eqref{eq:W-free}, we observe that $F[h]$ is given by a mean field like formula -- the first term $\int dx \log\big(1+g(x)e(x)\big)$ -- with effective local  density $g(x)$ self-consistently determined from the non-coincident cumulants  -- by coupling it to an external field $q(x)$ whose Boltzmann distribution is fixed by $F_0$.

We shall apply this result to give a new presentation of the known large deviation rate function in the classical SSEP.  Recall that SSEP is a stochastic model suited for describing transport and density fluctuations in many particle systems out of equilibrium. 
Its rate function, denoted $I_\mathrm{ssep}[n]$, governs the rare large density fluctuations in the sense that the probability that the SSEP density profile $\mathfrak{n}(x)$ approaches a given profile $n(x)$ away from the mean, most probable profile is exponentially small~:
\beq
\mathbb{P}\left[\mathfrak{n}(\cdot) \approx n(\cdot)\right] \asymp_{N\to\infty} e^{-N \,I_\mathrm{ssep}[n] } ,
\eeq
with $N$ the number of sites. A more precise definition and description shall be given in Section \ref{sec:SSEP}.

The derivation of the new formula we shall give uses three ingredients~: 
(i) First, the relation between SSEP and Q-SSEP~\cite{BernardJin19};
(ii) Second, the connections between the invariant measure of the quantum SSEP and free probability~\cite{Biane22};
(iii) Third, the solution of the problem stated above.

Combining these first two ingredients leads to a representation of the generating function for the non-coincident cumulants of the density in the classical SSEP in terms of appropriate free cumulants. Namely, let $F^\mathrm{ssep}_0[a]$ be the generating function of SSEP non-coincident cumulants, then
\[
F^\mathrm{ssep}_0[a] = \sum_{n\geq 1} \frac{(-1)^{n-1}}{n}R_n(\mathbb{I}_{[a]}),
\]
where the  $R_n$ are the free cumulants of the function $\mathbb{I}_{[a]}(x)=\int_x^1dy\, a(y)$ viewed as a random variable on the interval $[0,1]$ equipped with the Lebesgue measure as probability measure.

Knowing the generating function of the non-coincident cumulants, we can then use the solution \eqref{eq:F-bernoulli} of the problem stated above to write the large deviation rate function as the solution of the following extremization problem~:
\beq
{I}_\mathrm{ssep}[n]=\max_{g(\cdot), q(\cdot)} \left( \int_0^1\!\! dx \! \Big[ n(x)\log\big(\frac{n(x)}{g(x)}\big)+(1-n(x))\log\big(\frac{1-n(x)}{1-g(x)}\big)+q(x)g(x) \Big]-{F}^\mathrm{ssep}_0[q]\right).
\eeq
Comparing with the free formula \eqref{eq:I-free}, this formula has a mean field like self consistent flavor, as does the formula \eqref{eq:F-bernoulli}.
It also shows similarities with the formula known in the SSEP literature \cite{Derrida_Review,Mallick_Review} and we check in Section \ref{sec:equivalence} that they of course coincide. Its derivation is however different, as it makes a detour through Q-SSEP and it reveals the hidden ingredients from free probability in the classical SSEP large deviation rate function. 

Since the SSEP large deviation rate function has initially been derived using a matrix product ansatz for the SSEP stationary measure \cite{Derrida-etc,Derrida_Review,Mallick_Review}, one may wonder if there is any connection between matrix product ansatz, or more generally tensor network techniques, and free probability. In view of the impact of tensor techniques in studies of quantum many-body systems, such connection, if it exists, would provide further evidence for the possible universal role of free probability tools in such systems \cite{HruzaBernard22,Pappalardi22}.

The rest of the paper is organized as follows. In section 2 we show how to deal with cumulants of Bernoulli variables, using combinatorial techniques, and we derive the variational problem associated with the large $N$ limit. Another approach to these results, using more standard Feynman diagram tools is presented in section 3. Finally, in section 4, we make the connection with the Q-SSEP.

\section{Bernoulli Partition Functions and Combinatorics}
\label{sec:Combinatoire}
The purpose of this section is to give some combinatorial properties of cumulants, which will then be used to study the asymptotics of the free energy of a family of Bernoulli variables.
\subsection{Partition lattices and M\"obius functions}
\subsubsection{The lattice of partitions of a finite set}
 The set-partitions of   $\{1,\ldots,n\}$ (or, more generally, of a finite set $S$) form a lattice for the inverse refinement order, such that $\pi\leq \gamma$ if $\pi$ is finer than $\gamma$. We denote by   ${\mathcal P}_n$ (or ${\mathcal P}(S))$ this lattice. It has a  maximal element  $1_n$ (the partition with one part) and a minimal element $0_n$ (the partition with $n$ parts). Every interval $[\pi_1,\pi_2]$ in this lattice is isomorphic, as a partially ordered set, to a product
\begin{equation}\label{latticeiso}
[\pi_1,\pi_2]\sim \prod_{p}[0_{k_p},1_{k_p}]
\end{equation}
 where the  terms in the product are indexed by the parts $p$ of $\pi_2$ and $k_p$ is the number of parts of $\pi_1$ which are subsets of $p$.

\subsubsection{Lattice of partitions of a graph}
Let  $G$ be a finite, simple and loopless\footnote{There is a small terminology mismatch between communities here. In graph theory loopless means that there is no edge with its two ends at the same vertex. For Feynman graphs in physics, the term loop is used either for what is called a cycle in graph theory or a cycle class in homology, and this is the convention used in \autoref{sec:Feynman} and the Appendix. This should cause no confusion. Feynman graphs are neither simple --they may have multiple edges-- nor loopless in general. However they are for the situations covered in \autoref{sec:Feynman}.} graph (all graphs considered below will satisfy these conditions) with set of vertices  $V$ and edges $E$ and let ${\mathcal P}_G$ be the set of  partitions of $V$ into {\sl connected} parts.
Then  ${\mathcal P}_G\subset {\mathcal P}(V)$ with equality if and only if  $G$ is a complete   graph. We endow this set with the inverse refinement order $<_G$.

For every partition of $V$ there exists a maximal   partition $\pi^*\in {\mathcal P}_G$ such that $\pi^*\leq \pi$. The parts of this partition are the connected components of the parts of  $\pi$. It follows that
the partially ordered set ${\mathcal P}_G$ is a lattice with 
$$\pi_1\vee_G\pi_2=\pi_1\vee\pi_2\qquad\pi_1\wedge_G\pi_2=(\pi_1\wedge\pi_2)^*$$
Again there is a smallest element, $0_G$ and a maximal element $1_G$, whose parts are the connected components of $G$, moreover 
every interval  $[\pi_1,\pi_2]$ is isomorphic to a lattice of the form ${\mathcal P}_{G'}$ for some graph $G'$.
 
Every  partition $\pi\in {\mathcal P}_G$ defines a graph $G_\pi$ whose vertices are the parts of  $\pi$ and two vertices are connected by an edge if and only if  the union of the corresponding parts is   connected in $G$.
 In terms of the  graphs $G_\pi$ the covering relations for  the order on ${\mathcal P}_G$ can be described as the contraction of an edge:  $\pi_1<_G\pi_2$ is a covering relation if and only if $G_{\pi_2}$ can be obtained from $G_{\pi_1}$ by contracting some edge (an possibly removing spurious edges to keep the graph simple).
 
 For example, here is  ${\mathcal P}_G$ when $G$ is a cycle of size $4$. Each partition is denoted by its associated graph  $G_\pi$.  
 
 $$
\begin{tikzpicture}[scale=1]

  \draw [color=red](-3,2.6) -- (0,1.3)  -- (-1,2.6) ;\draw [color=red](3,2.6) -- (0,1.3)  -- (1,2.6) ;
  
    \draw [color=red](-3.5,4.4) -- (-5,5.6)  -- (-1,4.4) ;\draw [color=red](-3.5,4.4)  -- (-3,5.6)  -- (1,4.4) ;
    \draw [color=red](-3.5,4.4)  -- (-1,5.6)  -- (3.5,4.4) ;\draw [color=red] (-1,4.4) -- (1,5.6)  -- (1,4.4) ;\draw [color=red](-1,4.4) -- (3,5.6)  -- (3.5,4.4) ;
    \draw [color=red](1,4.4) -- (5,5.6)  -- (3.5,4.4) ;
 \draw [color=red](-5,7.4) -- (0,8.6)  -- (-3,7.4) ;\draw [color=red] (-1,7.4) -- (0,8.6)  -- (1,7.4) ;\draw [color=red](3,7.4) -- (0,8.6)  -- (5,7.4) ;

\draw (-.5,0) circle (.2cm);\draw (.5,0) circle (.2cm);
\node at (-.5,0){$\scriptstyle 4$};
\node at (.5,0) {$\scriptstyle 3$};
\draw (-.5,1) circle (.2cm);\draw (.5,1) circle (.2cm);
\node at (-.5,1){$\scriptstyle 1$};
\node at (.5,1) {$\scriptstyle 2$};
\draw(-0.5,.2)--(-0.5,.8);\draw(0.5,.2)--(0.5,.8);\draw(-0.3,0)--(.3,0);\draw(-0.3,1)--(.3,1);

\draw (-3.5,4) circle (.25cm);\draw (-2.5,3) circle (.2cm);
\node at (-3.5,4){$\scriptstyle 12$};
\node at (-2.5,3) {$\scriptstyle 3$};
\draw (-3.5,3) circle (.2cm);
\node at (-3.5,3){$\scriptstyle 4$};
\draw(-3.5,3.2)--(-3.5,3.75);\draw(-3.3,3)--(-2.7,3);\draw(-3.35,3.8)--(-2.65,3.1);

\draw (-.5,4) circle (.25cm);\draw (-1.5,4) circle (.2cm);
\node at (-.5,4){$\scriptstyle 23$};
\node at (-1.5,4) {$\scriptstyle 1$};
\draw (-1.5,3) circle (.2cm);
\node at (-1.5,3){$\scriptstyle 4$};
\draw(-1.3,4)--(-.75,4);\draw(-1.5,3.2)--(-1.5,3.8);\draw(-1.35,3.1)--(-.6,3.8);

\draw (1.5,3) circle (.25cm);\draw (.5,4) circle (.2cm);
\node at (1.5,3){$\scriptstyle 34$};
\node at (.5,4) {$\scriptstyle 1$};
\draw (1.5,4) circle (.2cm);
\node at (1.5,4){$\scriptstyle 2$};
\draw(.7,4)--(1.3,4);\draw(1.5,3.25)--(1.5,3.8);\draw(.6,3.85)--(1.35,3.2);

\draw (2.5,3) circle (.25cm);\draw (3.5,4) circle (.2cm);
\node at (2.5,3){$\scriptstyle 14$};
\node at (3.5,3) {$\scriptstyle 3$};
\draw (3.5,3) circle (.2cm);
\node at (3.5,4){$\scriptstyle 2$};
\draw(2.75,3)--(3.3,3);\draw(3.5,3.2)--(3.5,3.8);\draw(2.65,3.2)--(3.35,3.9);

\draw (-5,7) circle (.3cm);\draw (-5,6) circle (.2cm);
\node at (-5,7){$\scriptstyle 123$};
\node at (-5,6) {$\scriptstyle 4$};
\draw(-5,6.2)--(-5,6.7);

\draw (-3,7) circle (.25cm);\draw (-3,6) circle (.25cm);
\node at (-3,7){$\scriptstyle 12$};
\node at (-3,6) {$\scriptstyle 34$};
\draw(-3,6.25)--(-3,6.75);

\draw (-1,7) circle (.3cm);\draw (-1,6) circle (.2cm);
\node at (-1,7){$\scriptstyle 124$};
\node at (-1,6) {$\scriptstyle 3$};
\draw(-1,6.2)--(-1,6.7);

\draw (1,7) circle (.3cm);\draw (1,6) circle (.2cm);
\node at (1,7){$\scriptstyle 234$};
\node at (1,6) {$\scriptstyle 1$};
\draw(1,6.2)--(1,6.7);

\draw (3,7) circle (.25cm);\draw (3,6) circle (.25cm);
\node at (3,7){$\scriptstyle 14$};
\node at (3,6) {$\scriptstyle 23$};
\draw(3,6.25)--(3,6.75);

\draw (5,7) circle (.3cm);\draw (5,6) circle (.2cm);
\node at (5,7){$\scriptstyle 134$};
\node at (5,6) {$\scriptstyle 2$};
\draw(5,6.2)--(5,6.7);

\draw (0,9) circle (.35cm);\node at (0,9){$\scriptstyle 1234$};
\end{tikzpicture}
$$

\subsubsection{M\"obius functions}
Recall that, for a partially ordered set,  its {\sl zeta function} is the function
$$
\begin{array}{rcl}
 \zeta(x,y)&=&1\quad \text{if } x\leq y
\\
&=& 0\quad\text{if not} 
\end{array}
$$
The {\sl M\"obius function} $\mu(x,y)$, defined for $x\leq y$, satisfies, for all $x\leq z$:
$$\sum_{y;x\leq y\leq z}\mu(x,y)\zeta(y,z)=\delta_{xz}$$

The M\"obius functions for the lattices $\mathcal{P}_n$ and, more generally, $\mathcal{P}_G$ play an important role in the following.
The  M\"obius  function on $\mathcal{P}_n$ is  multiplicative namely if 
$[\pi_1,\pi_2]$ is as in (\ref{latticeiso}) then 
$$\mu(\pi_1,\pi_2)=\prod_p\mu(0_{k_p},1_{k_p})$$
  and   $$\mu(0_k,1_k)=(-1)^{k-1}(k-1)!$$ In order to compute the M\"obius function on $\mathcal{P}_G$ we will need some facts about chromatic polynomials.
\subsubsection{Chromatic polynomials}
A {\sl proper colouring} of a finite graph  $G$ is a colouring of its vertices such that, for any edge, the  adjacent vertices have different colours.
The chromatic polynomial of $G$, denoted $\chi_G$, is the unique polynomial such that, for any integer $k\geq 1$ the number of proper colourings of $G$ with at most  $k$ colours  is equal to $\chi_G(k)$. If $\omega_r$ denotes the number of proper colourings of $G$ which use exactly $r$ colours then one has
\begin{equation}\label{color}
 \chi_G(k)=\sum_r\omega_r{k\choose r}.
 \end{equation}
Since $\omega_r=0$ for $r>|V|$ this shows that $\chi_G$ is indeed a polynomial.

For example, the complete graph with  $n$ vertices has  $$\chi_{K_n}(z)=(z)_n:=z(z-1)(z-2)\ldots (z-n+1)$$ while, if $T$ is a tree with $n$ vertices, then  
$$\chi_{T}(z)=z(z-1)^{n-1}.$$

 We note the following properties of the chromatic polynomial:
 if $G$ is the union of two disjoint graphs $G_1,G_2$ then  
 \begin{equation}\label{disjoint}
 \chi_G(z)=\chi_{G_1}(z)\chi_{G_2}(z)
\end{equation}
 whereas, if $G$ is the join of $G_1,G_2$, namely $V=V_1\cup V_2$ and $V_1\cap V_2=\{v\}$ with no edge joining $V_1\setminus \{v\}$ to $V_2\setminus\{v\}$ then
\begin{equation}\label{join}
 \chi_G(z)=\frac{1}{z}\chi_{G_1}(z)\chi_{G_2}(z).
\end{equation}

The M\"obius function of ${\mathcal P}_G$ has been computed by Rota \cite{R}, one has
\begin{equation}\label{Moeb}\mu(0_G,1_G)=[z]\chi_G(z),\end{equation}
the coefficient of $z$ in the polynomial $\chi_G(z)$,
moreover, if  $\pi_1\leq\pi_2$ in ${\mathcal P}_G$ then $[\pi_1,\pi_2]\sim {\mathcal P}_{G'}$ for some graph $G'$ and
$$\mu(\pi_1,\pi_2)=\mu(0_{G'},1_{G'})$$
Note that, by (\ref{color}), one has 
\begin{equation}\label{color2}
[z]\chi_G(z)=\sum_r\frac{(-1)^{r-1}}{r}\omega_r
\end{equation}
In the following we will use the notation $\mu(G)=\mu(0_G,1_G)$ when the context is clear.

The proof of (\ref{Moeb}) is based on inclusion-exclusion. The number of all colourings of $G$ using at most $k$ colours is $k^{|V|}$, moreover any such colouring determines a partition $\pi\in{\mathcal P}_G$ into connected unicolour components, so that the graph $G_\pi$ is properly coloured. 
It follows that
$$k^{|V|}=\sum_{\pi\in {\mathcal P}_G}\chi_{G_\pi}(k)$$
and formula (\ref{Moeb}) is obtained  by M\"obius inversion, see \cite{R} for details.

\subsection{Moments and cumulants}
Let  ${\mathcal A}$ be a complex  algebra with unit and   $\varphi:{\mathcal A}\to {\bf C}$ a linear form such that $\varphi(1)=1$.
For most applications below  ${\mathcal A}$ will be an algebra of  complex random variables defined over some probability space, in particular it will be commutative, but it is not more difficult to consider here the  general case of an arbitrary algebra over the complex numbers.

The {\sl cumulants} are a sequence of $n$-multilinear forms  $K_n, n=1,2,\ldots$ on ${\mathcal A}$, implicitely defined by 
\begin{equation}\label{mom-cum}
\varphi(a_1\ldots a_n)=\sum_{\pi\in {\mathcal P}_n} K_\pi(a_1,\ldots,a_n)
\end{equation} 
with 
\begin{equation}\label{mom-cum1}
K_\pi(a_1,\ldots,a_n)=\prod_{p\in \pi}K_{|p|}(a_{i_1},\ldots,a_{i_{|p|}})
\end{equation} 
the product being over the parts of $\pi$ with 
 $p=\{i_1,\ldots,i_{|p|}\}$ and $i_1<i_2<\ldots<i_{|p|}$.
This formula can be inverted to express the cumulants in terms of the ``moments'', i.e. $\varphi$ evaluated on products.
For example  
\begin{eqnarray*}
\varphi(a_1) &=& K_1(a_1) \\
\varphi(a_1a_2) &=& K_2(a_1,a_2)+K_1(a_1)K_1(a_2)
\end{eqnarray*}
 gives $$K_2(a_1,a_2)=\varphi(a_1a_2)-\varphi(a_1)\varphi(a_2)$$
while 
$$\begin{array}{rcl}\varphi(a_1a_2a_3)&=&K_3(a_1,a_2,a_3)+
 K_1(a_1)K_2(a_2,a_3)+
 K_2(a_1,a_3)K_1(a_2)\\&&
 +K_2(a_1,a_2)K_1(a_3)+
 K_1(a_1)K_1(a_2)K_1(a_3)\end{array}$$
gives
$$\begin{array}{rcl}
K_3(a_1,a_2,a_3)&=&\varphi(a_1a_2a_3)-\varphi(a_1a_2)\varphi(a_3)-\varphi(a_1a_3)\varphi(a_2)\\&&
-\varphi(a_1)\varphi(a_2a_3)
+2\varphi(a_1)\varphi(a_2)\varphi(a_3)\end{array}$$

In the general case there is an expression using the M\"obius function on $ {\mathcal P}_n$:

$$K_n(a_1,\ldots,a_n)=\sum_{\pi\in {\mathcal P}_n} \varphi_\pi(a_1,\ldots, a_n)\mu(\pi,1_n)$$

In the case where ${\mathcal A}$ is commutative the cumulants are symmetric multilinear forms and their generating function is
\begin{equation}\label{free-energy0}
\log \varphi[e^{\sum_{i=1}^N\lambda_ia_i}]=\sum_{n=1}^\infty\frac{1}{n!}\sum_{I:i_1+\ldots+i_N=n}\lambda_1^{i_1}\ldots\lambda_N^{i_N}K_n(a_I)
\end{equation}
where one sums over all sequences  $a_I=(a_{j_1},\ldots, a_{j_n})$ with $i_k$ occurrences of $a_k$.
Each such sequence determines a partition of $\{1,\ldots,n\}$ into parts corresponding to the value of the indices.
One can thus rewrite (\ref{free-energy0}) as
\begin{equation}\label{free-energy}
\log \varphi[e^{\sum_{i=1}^N\lambda_ia_i}]=\sum_{n=1}^\infty\frac{1}{n!}\sum_{\Gamma\in{\mathcal LP}_n}\lambda^\Gamma K_n(a_\Gamma)
\end{equation}
where the sum is over labelled partitions   $\Gamma$ of $\{1,\ldots,N\}$ into at most $N$ parts, where each part $\gamma$ of $\Gamma$
 has a label $\nu(\gamma)$ in $\{1,2,\ldots,N\}$ (the parts having distinct labels) and $\lambda^\Gamma=\prod_{\gamma\in \Gamma}\lambda_{\nu(\gamma)}^{|\gamma|}$.
  
\subsubsection{Cumulants with products as entries}\label{prod_entr}

Let  $\Gamma:=\gamma_1\cup\ldots\cup\gamma_k$ be a partition of $\{1,\ldots,n\}$ into intervals i.e. each $\gamma_l$ is of the form
$\{j_l+1,j_l+2,\ldots,j_{l+1}\}$ with $0=j_1<j_2<\ldots <j_{k+1}=n$.

Let us define
 $$K_n^{\Gamma}(a_1,\ldots,a_n)=K_k(A_1,\ldots,A_k)$$ 
where
$A_l=a_{j_l+1}a_{j_l+2}\ldots a_{j_{l+1}}$, the product of the $a_i$ with indices $i\in\gamma_l$ and, more generally,
\begin{equation}\label{gamm-cum}
K^\Gamma_\pi(a_1,\ldots,a_n)=\prod_{p\in\pi}K_{|p|}^{\Gamma_{|p}}(a_{i_1},\ldots,a_{i_{|p|}})
\end{equation} 
Here  $\Gamma_{|p}$ is the partition of $p\in\pi$ induced by $\Gamma$. Observe that one has also
\begin{equation}\label{gamm-cum2}
K^\Gamma_\pi(a_1,\ldots,a_n)=K_\pi^{\Gamma\wedge \pi}(a_1,\ldots,a_n).
\end{equation} 
One has 
$$
\begin{array}{rcl}
K^{0_n}_\pi(a_1,\ldots,a_n)&=&K_\pi(a_1,\ldots,a_n)\\
K^{1_n}_\pi(a_1,\ldots,a_n)&=&\varphi_\pi(a_1\ldots a_n)
\end{array}
$$
 so that the  $K^\Gamma_\pi$, for 
$0_n\leq\Gamma\leq 1_n$, interpolate between cumulants and moments. The following formula, attributed to Leonov and Shiryaev \cite{LS}, expresses the $K^\Gamma$ in terms of ordinary cumulants.

\begin{thm}
\begin{equation}\label{cum-prod}
K^{\Gamma}_\xi(a_1,\ldots,a_n)=\sum_{\pi:\pi\vee\Gamma=\xi}K_\pi(a_1,\ldots,a_n),\qquad \text{for}\quad \xi\geq \Gamma . 
\end{equation}
In particular 
\begin{equation}\label{cum-prod1}
K_n^{\Gamma}(a_1,\ldots,a_n)=\sum_{\pi:\pi\vee\Gamma=1_n}K_\pi(a_1,\ldots,a_n) .
\end{equation}
\end{thm}

\begin{proof} This follows easily by comparing the two formulas:
$$
\begin{array}{rcl}
\varphi(a_1\ldots a_n)&=&\sum_{\pi}K_\pi(a_1,\ldots,a_n)=\sum_{\xi\geq \Gamma}\left(\sum_{\pi:\pi\vee\Gamma=\xi}K_\pi(a_1,\ldots,a_n)\right)\\
\varphi(a_1\ldots a_n)&=&\varphi(A_1\ldots A_n)=\sum_{\xi\geq \Gamma}K_\xi^\Gamma(a_1,\ldots,a_n)
\end{array}
$$
\end{proof}
When  $\Gamma=0_n$ the formula (\ref{cum-prod1}) is trivially true and when $\Gamma=1_n$ it is the moments-cumulants formula (\ref{mom-cum}).
Also if $\xi\ngeq\Gamma$ one can use (\ref{gamm-cum2}) to get
\begin{equation}\label{cum-prod2}
K^{\Gamma}_\xi(a_1,\ldots,a_n)=K^{\Gamma\wedge\xi}_\xi(a_1,\ldots,a_n)=\sum_{\pi:\pi\vee(\Gamma\wedge\xi)=\xi}K_\pi(a_1,\ldots,a_n)
\end{equation}
In the general case the formula (\ref{cum-prod1}) can be inverted. For this we introduce a graph
 $G_{\pi,\Gamma}$,
 whose vertices are the parts of $\mu$, and there is an edge between  $p$ and $q$ if there exists a part $\gamma$ of $\Gamma$ such that
 $p\cap \gamma\ne\emptyset$ and $q\cap \gamma\ne\emptyset$.
One has then
\begin{thm}
\begin{equation}
\label{gamm-cuminv}
K_n(a_1,\ldots,a_n)=\sum_{\pi:\pi\vee\Gamma=1_n}K_\pi^\Gamma(a_1,\ldots,a_n)\mu(G_{\pi,\Gamma})
\end{equation}
\end{thm}
\begin{proof}
This formula can be verified by plugging it into the right hand side of (\ref{cum-prod}) and checking that it reduces to $K_n^{\Gamma}(a_1,\ldots,a_n)=K_n^{\Gamma}(a_1,\ldots,a_n)$ after using the properties of the M\"obius function.

Introducing the  unknown function $\mu_\Gamma(\pi)$ such that 
$$K_n(a_1,\ldots,a_n)=\sum_{\pi:\pi\vee\Gamma=1_n}K_\pi^\Gamma(a_1,\ldots,a_n)\mu_\Gamma(\pi)$$
and using (\ref{cum-prod2}) one has 
$$
\begin{array}{rcl}
K_n(a_1,\ldots,a_n)&=&\sum_{\pi:\pi\vee\Gamma=1_n}K_\pi^\Gamma(a_1,\ldots,a_n)\mu_\Gamma(\pi)\\
&=&\sum_{\pi:\pi\vee\Gamma=1_n}\sum_{\xi:(\pi\wedge \Gamma)\vee\xi=\pi}K_\xi(a_1,\ldots,a_n)\mu_\Gamma(\pi)\\
&=&\sum_{\xi:\xi\vee\Gamma=1_n}K_\xi(a_1,\ldots,a_n)\left(\sum_{\pi:(\pi\wedge \Gamma)\vee\xi=\pi}\mu_\Gamma(\pi)\right)\\
\end{array}
$$
so that $\mu_\Gamma$
 has to satisfy

\begin{equation}
\label{moeb}
\sum_{\pi:(\pi\wedge \Gamma)\vee\xi=\pi}\mu_\Gamma(\pi)=1,\quad \text{if}\quad \xi=1_n, \quad
=0\quad \text{if not.}
\end{equation}

Define an order relation on the set of partitions $\pi$ such that $\pi\vee\Gamma=1_n$ by requiring
$$\pi_1\leq_\Gamma\pi_2\quad\text{ if and only if}\quad \pi_1\leq\pi_2\quad\text{and}\quad \pi_1\vee(\pi_2\wedge \Gamma)=\pi_2$$
Indeed, if  $\pi_1\leq_{\Gamma}\pi_2$ and  $\pi_2\leq_{\Gamma}\pi_3$ then $\pi_1\vee(\pi_3\wedge\Gamma)\geq \pi_1\vee(\pi_2\wedge\Gamma)=\pi_2$ therefore
$\pi_1\vee(\pi_3\wedge\Gamma)\geq \pi_2$ and $\pi_1\vee(\pi_3\wedge\Gamma)\geq \pi_3\wedge\Gamma$ so that, finally $\pi_1\vee(\pi_3\wedge\Gamma)\geq \pi_2\vee(\pi_3\wedge\Gamma)=\pi_3$. This relation is transitive as claimed. 
Taking $\mu_\Gamma(\pi)=\mu([\pi,1_n])$ where $\mu$ is the M\"obius function for this order yields  (\ref{moeb}).
It is easy to check that the interval $[\pi,1_n]$ for this order is   isomorphic \`a ${\mathcal P}_{G_{\pi,\Gamma}}$.
One has thus $\mu_\Gamma(\pi)=[z]\chi_{G_{\pi,\Gamma}}=\mu(G_{\pi,\Gamma})$. 

\end{proof}
In the commutative case, one can define the cumulants $K^{\Gamma}$ for any partition $\Gamma$, not just interval partitions, since the cumulants are symmetric and (\ref{cum-prod}),  (\ref{cum-prod1}),
(\ref{gamm-cuminv}) still hold.
\subsection{Non-crossing partitions and cumulants}
In this section we define non-crossing cumulants which will be used later in the asymptotic analysis of the QSSEP. Many informations about the combinatorics of the non-crossing cumulants may be found in the book by Nica and Speicher \cite{NS}.

A partition of $\{1,\ldots,n\}$ has a crossing if there exists two parts of the partition and $i<j<k<l$ such that $i,k$ belong to the first part and $j,l$ to the second part. Partitions without crossing are called {\sl non-crossing}. The set of non-crossing partitions of $\{1,\ldots,n\}$, denoted $NC(n)$,  is a lattice under the inverse refinement order, and each interval $[\pi_1,\pi_2]$ is isomorphic, as a partially ordered set, to a product
$\prod_iNC(k_i)$ for some integers $k_i$. The M\"obius function is again multiplicative and one has
$$\mu_{NC(n)}(0_n,1_n)=(-1)^{n-1}\text{Cat}_{n-1}$$ where $\text{Cat}_n=\frac{1}{n+1}{2n\choose n}$ is a Catalan number.

Non-crossing cumulants $R_n$ are defined similarly as the cumulants $K_n$ using an implicit formula:
\begin{equation}\label{free-mom-cum}
\varphi(a_1\ldots a_n)=\sum_{\pi\in NC(n)} R_\pi(a_1,\ldots,a_n)
\end{equation} 
which can be inverted as
$$R_n(a_1,\ldots,a_n)=\sum_{\pi\in NC(n)} \varphi_\pi(a_1,\ldots, a_n)\mu_{NC(n)}(\pi,1_n)$$
Every partition $\pi\in{\mathcal P}_n$ has a least non-crossing majorant $\hat\pi$. Using this one can write

$$
\begin{array}{rcl}
\varphi(a_1\ldots a_n)&=&\sum_{\pi\in{\mathcal P}_n }K_\pi(a_1,\ldots,a_n)=\sum_{\xi\in NC(n)}\left(\sum_{\hat\pi=\xi}K_\pi(a_1,\ldots,a_n)\right)\\
\varphi(a_1\ldots a_n)&=&\sum_{\xi\in NC(n)}R_\xi(a_1,\ldots,a_n)
\end{array}
$$
from which one can easily deduce that 
$$R_\xi(a_1,\ldots,a_n)=\sum_{\pi:\hat\pi=\xi}K_\pi(a_1,\ldots,a_n) .$$
In particular, the relation
\begin{equation}
R_n(a_1,\ldots,a_n)=\sum_{\pi:\hat\pi=1_n}K_\pi(a_1,\ldots,a_n)
\end{equation}
 expresses non-crossing cumulants in terms of cumulants.
The formula can be reversed using again the M\"obius function of a certain lattice ${\mathcal P}_G$. For this, define the {\sl crossing graph} $G^c_\pi$ of a partition $\pi$ as the graph whose vertices are the parts of $\pi$ and two parts of $\pi$ are connected if they contain  a crossing.
Using this graph one has, by M\"obius inversion,
\begin{prop}
\begin{equation}
K_n(a_1,\ldots,a_n)=\sum_{\pi:\hat\pi=\xi}R_\pi(a_1,\ldots,a_n)\mu(G^c_\pi).
\end{equation}
\end{prop}
An equivalent formula was first derived in \cite{MJV} by different means, see also \cite{AHLV}.

There is also, for non-crossing cumulants, an analogue of the formula  (\ref{cum-prod1}), due to Krawczyk and Speicher \cite{KS}.
\subsection{Cumulants of Bernoulli variables}\label{ssec:CBv}
\subsubsection{Non-coincident cumulants }
Let  $b_i; i=1,2,\ldots,N$ be a sequence of (commuting) Bernoulli random variables, taking values in $\{0,1\}$. They  satisfy $b_i^r=b_i$ for all $r\geq 1$ therefore
all the information about the joint distribution of the $b_i$ is contained in the $2^N-1$ ``non-coincident moments'', i.e. the quantities
 $E[b_{i_1}b_{i_2}\ldots b_{i_k}]$ where $1\leq i_1<i_2<\ldots <i_k\leq N$ (here $E$ denotes the expectation), or  in the $2^N-1$ ``non-coincident  cumulants'' 
$K_k(b_{i_1},b_{i_2},\ldots,b_{i_k})$. It is therefore of interest to express an arbitrary cumulant 
$K_n(b_{j_1},\ldots,b_{j_n})$, for a sequence of indices $1\leq j_k\leq N$, in terms of these non-coincident  cumulants.

Let  $\Gamma$ be the partition of  $\{1,\ldots,n\}$ such that  $k$ and $l$ are in the same part of $\Gamma$ if and only if  $i_k=i_l$. Using the fact that $(b_i)^r=b_i$ for any $r\geq 1$ we see that the $\Gamma$-cumulants defined by (\ref{gamm-cum}) and the formula (\ref{gamm-cuminv}) express any cumulant as a polynomial in  the non-coincident  cumulants. 

\begin{equation}
\label{gamm-cuminv1}
K_n(b_{i_1},\ldots,b_{i_n})=\sum_{\pi:\pi\vee\Gamma=1_n}K_\pi^\Gamma(b_{i_1},\ldots,b_{i_n})\mu(G_{\pi,\Gamma})
\end{equation}

\subsubsection{Free energy} 
Recall the generating function of the cumulants (\ref{free-energy})
\begin{equation}
\log E[e^{\sum_{i=1}^Nh_ib_i}]=\sum_{n=1}^\infty\frac{1}{n!}\sum_{\Gamma}h^\Gamma K_n(b_\Gamma)
\end{equation}
One can use  (\ref{gamm-cuminv1}) in each term of this sum  to obtain a sum over pairs $(\pi,\Gamma)$:

\begin{equation}\label{free-energy-pigamma}
\log E[e^{\sum_{i=1}^Nh_ib_i}]=\sum_n\frac{1}{n!}\sum_\Gamma h^\Gamma
\left(\sum_{\pi:\pi\vee\Gamma=1_n} K_\pi^\Gamma(b_\Gamma)\mu(G_{\pi,\Gamma})\right).
\end{equation}
Let us introduce a labelled bipartite graph $\Delta_{\pi,\Gamma}$ with  the parts of  $\Gamma$  as  set of white vertices and the parts of  $\pi$ as  set of black vertices. The white vertices, corresponding to the parts of $\Gamma$, are labelled by   $1,2,\ldots$ (the indices of the Bernoulli variables) each index appearing at most once. There is an edge between a part of $\Gamma$ and a part of $\pi$ if they have a non-empty intersection. 
The condition $\pi\vee\Gamma=1_n$ ensures that this graph is connected. Observe that one can associate to every edge of the graph a subset of $\{1,\ldots,N\}$ by taking the intersection of the part of $\pi$ corresponding to its black extremity and the part of $\Gamma$ corresponding to its white extremity. These sets form a partition of $\{1,\ldots,N\}$, indexed by the edges of the graph,  and one can reconstruct   the partitions $\pi$ and $\Gamma$ from this edge-indexed partition by taking the union over edges adjacent to a white vertex to get the parts of  $\Gamma$ or to a black vertex, to get the parts of $\pi$.

 The graph  $G_{\pi,\Gamma}$ is obtained from the bipartite graph $\Delta_{\pi,\Gamma}$ by keeping the black vertices and putting an edge between two such vertices if they have at least one  white neighbour in common.

The factor associated with the pair $\pi,\Gamma$ can then be written as 
 \begin{equation}\label{factor}
 \mu(G_{\pi,\Gamma})\frac{h^\Gamma}{n!}\prod_{\bullet}K(b_\bullet)
 \end{equation} 
where the product is over the black vertices $\bullet$ of $\Delta_{\pi,\Gamma}$ and, for each such vertex, the  factor $$K(b_\bullet)=
K_k(b_{u_1},b_ {u_2},\ldots,b_{u_k})$$
 the   indices $u_1,u_2,\ldots, u_k$ being those of the white neighbours of $\bullet$ in $\Delta_{\pi,\Gamma}$.

 As an example, here is the graph $\Delta_{\pi,\Gamma}$ associated with the partitions $$\pi=\{1,2,5,8,14\}\cup\{3,6,10,12\}\cup\{4,7,9,11,113\}$$ and $$\Gamma=\{1,5,13,14\}\cup\{2,6,8,9,11\}\cup\{3,4,7\}\cup\{10,12\}$$
 where we show, near each edge, the associated set:
 
$$
\begin{tikzpicture}[scale=1]
\draw[fill] (0,0) circle (.4cm);\draw[color=blue] (2,0) circle (.4cm);\draw[fill] (2,2) circle (.4cm);\draw[color=blue] (0,2) circle (.4cm);\draw[color=blue] (2,4) circle (.4cm);
\draw[fill] (-1,3) circle (.4cm);\draw[color=blue] (-1,-1) circle (.4cm);
\node[color=red] at (.5,3.8){$\scriptstyle \{1,5,14\}$};\node[color=red] at (-.9,2.3){$\scriptstyle \{2,8\}$};
\node[color=red] at (1,2.2){$\scriptstyle \{9,11\}$};\node[color=red] at (-.4,1){$\scriptstyle \{6\}$};\node [color=red] at (2.4,1){$\scriptstyle \{4,7\}$};
\node[color=red] at (1,.2){$\scriptstyle \{3\}$};\node[color=red] at (.1,-.65){$\scriptstyle \{10,12\}$};\node[color=red] at (2.4,3){$\scriptstyle \{13\}$};
\draw[color=red] (0,.4)--(0,1.6);\draw[color=red] (2,.4)--(2,1.6);\draw[color=red] (0.4,0)--(1.6,0);\draw[color=red] (-0.7,-0.7)--(-0.3,-0.3);\draw[color=red] (2,.4)--(2,1.6);\draw[color=red] (0.4,2)--(1.6,2);
\draw[color=red] (-0.7,2.7)--(-0.3,2.3);\draw[color=red] (-0.6,3)--(1.6,3.85);\draw[color=red] (2,2.4)--(2,3.6);

\end{tikzpicture}
$$ 
 
Let $H$ be a bipartite connected graph (with at least two vertices), then the pairs $(\pi,\Gamma)$ such that $H$ is the underlying unlabelled graph of $\Delta_{\pi,\Gamma}$ can be obtained by 
\begin{enumerate}
\item Labelling the white vertices of $H$ with distinct labels in $\{1,2,\ldots,N\}$ (we call $Lab(H)$ the set of such labellings).
\item
 Choosing a partition of $\{1,\ldots,N\}$ indexed by the edges of $H$.
\end{enumerate}
Denote  $H^\bullet$ the graph whose vertices are the  black vertices of $H$ and with edges between vertices sharing a white neighbour in $H$.
Using the fact that we are summing over partitions indexed by edges of $H$, which are counted by multinomial coefficients,  we see that
the sum of all contributions (\ref{factor}) corresponding to $H$ is 
\begin{equation} \label{eq:weight-L0}
  \frac{\mu (H^\bullet)}{|\text{Aut}\, H|}\sum_{{\mathcal L}\in Lab(H)}w({\mathcal L})
\end{equation}
where  ${\mathcal L}$ runs over all labellings of the white vertices of $H$ by distinct indices and
\begin{equation} \label{eq:weight-L}
w({\mathcal L})=\prod_{\bullet}K(b_\bullet)\prod_{\text{edges of $H$}}e_i
\end{equation}
 (for any edge $e$ of $H$ one denotes $e_i:=e^{h_i}-1$, where $i$ is the index of the white vertex adjacent to the edge $e$). The term $\frac{1}{|\text{Aut}\, H|}$, as usual, is here to avoid overcountings due to symmetries. The automorphism group is that of $H$ considered as a bipartite graph, i.e. automorphisms should send black vertices to black vertices and white vertices to white vertices.
 We can thus rewrite (\ref{free-energy-pigamma}) as

\begin{prop}
\begin{equation}\label{Weq}
W[h]=\log E\big[e^{\sum_{i=1}^Nh_ib_i}\big]=\sum_H\frac{\mu (H^\bullet)}{|\text{Aut}\, H|}\sum_{{\mathcal L}\in Lab(H)}w({\mathcal L}) ,
\end{equation}
where the sum is over connected bipartite graphs and the weight $w({\mathcal L})$ as in eq.\eqref{eq:weight-L}.
\end{prop}

Here the  graph $H$ corresponding to the pair $\pi,\Gamma$ above, with a labelling  by  $1,2,5,6$. 
$$
\begin{tikzpicture}[scale=1]
\draw[fill] (0,0) circle (.4cm);\draw[color=blue] (2,0) circle (.4cm);\draw[fill] (2,2) circle (.4cm);\draw[color=blue] (0,2) circle (.4cm);\draw[color=blue] (2,4) circle (.4cm);
\draw[fill] (-1,3) circle (.4cm);\draw[color=blue] (-1,-1) circle (.4cm);
\node[color=blue] at (2,0){$1$};\node[color=blue] at (0,2){$2$};\node[color=blue] at (-1,-1){$5$};\node[color=blue] at (2,4){$6$};
\node at (-2.3,3){$\scriptstyle\bf K_2(b_2,b_6)$};
\node at (3.6,2){$\scriptstyle \bf K_3(b_1,b_2,b_6)$};\node at (-1.4,.4){$\scriptstyle\bf K_3(b_1,b_2,b_5)$};
\node[color=red] at (.5,3.8){$\scriptstyle e_6$};\node[color=red] at (-.7,2.3){$\scriptstyle e_2$};
\node[color=red] at (1,2.2){$\scriptstyle e_2$};\node[color=red] at (-.4,1){$\scriptstyle e_2$};\node [color=red] at (2.4,1){$\scriptstyle e_1$};
\node[color=red] at (1,.2){$\scriptstyle e_1$};\node[color=red] at (-.25,-.65){$\scriptstyle e_5$};\node[color=red] at (2.4,3){$\scriptstyle e_6$};
\draw[color=red] (0,.4)--(0,1.6);\draw[color=red] (2,.4)--(2,1.6);\draw[color=red] (0.4,0)--(1.6,0);\draw[color=red] (-0.7,-0.7)--(-0.3,-0.3);\draw[color=red] (2,.4)--(2,1.6);\draw[color=red] (0.4,2)--(1.6,2);
\draw[color=red] (-0.7,2.7)--(-0.3,2.3);\draw[color=red] (-0.6,3)--(1.6,3.85);\draw[color=red] (2,2.4)--(2,3.6);

\end{tikzpicture}
$$
The graph  $H^\bullet$ is a complete graph with three vertices so that  $\mu(H^\bullet)=2$ and there are no nontrivial automorphisms moreover the  weight of the labelling is 
$$w({\mathcal L})=e_1^2e_2^3e_5e_6^2K_3(b_1,b_2,b_6)K_2(b_2,b_6)K_3(b_1,b_2,b_5)$$

\subsubsection{Another proof of formula (\ref{Weq})}
We sketch another derivation of (\ref{Weq}), which does not rely on the theory of cumulants of products.
One has
$$e^{\sum_{i=1}^Nh_ib_i}=\prod_i(1+e_ib_i)=1+\sum_{I\subset \{1,\ldots,N\};I\ne \emptyset}e_Ib_I$$
where $e_I$ is the product $\prod_{i\in I}e_i$. Using the moment-cumulant formula we get
\begin{equation}
E[e^{\sum_{i=1}^Nh_ib_i}]=1+\sum_{I\ne \emptyset}e_I\sum_{\pi\in{\mathcal P}(I)}K_\pi(b_I)
\end{equation}
and taking the logarithm
$$W[h]=\sum_{r=1}^\infty \frac{(-1)^r}{r}\left(\sum_{I\ne \emptyset}e_I\sum_{\pi\in{\mathcal P}(I)}K_\pi(b_I)\right)^r$$
One has
$$\left(\sum_{I\ne \emptyset}e_I\sum_{\pi\in{\mathcal P}(I)}K_\pi(b_I)\right)^r=\sum_{I_1,\ldots,I_r,\pi_1\ldots,\pi_r}\prod_{k=1}^rK_{\pi_k}(b_{I_k})e_{I_k}$$
where we sum over $I_1\ldots, I_r$, non-empty subsets of  $[1,N]$ and $\pi_k$  partition of  $I_k$.

We  now introduce a bipartite graph with white vertices labelled by the $i\in\cup_k I_k$ and black vertices corresponding to the parts of the partitions $\pi_k$. There is an edge between a white and a black vertex if the index of the white vertex is in the part corresponding to the black vertex. 
This bipartite graph induces a graph structure on the black vertices: two vertices share an edge if they have a common white neighbour.
For each such graph we have to sum over all proper colourings of the black vertices using exactly $r$ colours. Using relation 
 (\ref{color}) we identify the combinatorial term associated with a graph to the   $z$ coefficient  in the chromatic polynomial of the black graph.
 By (\ref{disjoint}) this coefficient is zero is the graph is not connected so that the sum can be taken over connected graph We leave details to the reader and give an example: it is easy to see that the  the monomial $$e_1^3e_2^3K_1(b_1)^2K_1(b_2)^2K_2(b_1,b_2)$$ is obtained from only one graph $H$, the one depicted below.
 
\begin{equation} \label{eq:graphil}
\begin{tikzpicture}[scale=1]
\draw[fill] (0,0) circle (.4cm);\draw[color=blue] (2,0) circle (.4cm);\draw[fill] (3,1) circle (.4cm);\draw[color=blue] (-2,0) circle (.4cm);
\draw[fill] (3,-1) circle (.4cm);\draw[fill] (-3,-1) circle (.4cm);\draw[fill] (-3,1) circle (.4cm);
\draw(-2.7,.7)--(-2.3,.3);\draw(-2.7,-.7)--(-2.3,-.3);\draw(2.7,.7)--(2.3,.3);\draw(2.7,-.7)--(2.3,-.3);
\draw(-1.6,0)--(-.4,0);\draw(1.6,0)--(.4,0);
\end{tikzpicture}
\end{equation}

The graph $H^\bullet$ is the join of two complete graphs therefore $\mu(H^\bullet)=4$ while $|\text{Aut}(H)|=8$, moreover there are two labellings of the white vertices by $1,2$ therefore the sum of coefficients  of this graph is $1$, which should be the coefficient of the monomial. 

On the other hand one can obtain the coefficient of this monomial  by expanding the expression 
$\frac{1}{3}w^3-\frac{1}{4}w^4+\frac{1}{5}w^5$ (other powers of $w$ do not contribute) where $$w=e_1K_1(b_1)+e_2K_1(b_2)+e_1e_2K_2(b_1,b_2)+e_1e_2K_1(b_1)K_1(b_2)$$
 Using multinomial coefficients we find
$$\frac{1}{3}\frac{3!}{2!1!}-\frac{1}{4}\frac{4!}{1!1!1!}+\frac{1}{5}\frac{5!}{1!2!2!}=1,$$ 
so that the weight of this graph is effectively $1$.

\subsection{Asymptotic behaviour and Legendre transform}
\subsubsection{Reduction to trees}
We suppose now that, as  $N\to\infty$, the non-coincident cumulants  have a specific asymptotic behaviour: there exists some compact space $\Sigma$, some functions $\rho_N:[1,N]\to \Sigma$ and continuous functions 
  $\psi_n$ on $\Sigma^n$ such that as $N\to\infty$
\begin{equation}\label{order}
 K_n(b_{i_1},\ldots,b_{i_n})\sim N^{1-n}\psi_n(\rho_N(i_1),\ldots,\rho_N(i_n))
\end{equation}
Moreover the measures
$\frac{1}{N}\sum_i\delta_{\rho_N(i)}$ converge to some diffuse measure $ds$ on $\Sigma$.
We are mainly interested in the case where $\rho_N(i)=i/N$ and $\Sigma=[0,1]$ but the analysis works in greater generality and can be adapted to deal with other topologies, e.g. lattices in higher dimension.

We will study the asymptotic behaviour of the free energy and for this 
we assume that the $h_i$ converge also to some bounded function $h(s)$ on $\Sigma$. We  can then estimate the contribution of a graph $H$ in (\ref{Weq}). Indeed the number of labellings is
 $$(N)_{\sharp\text{white vertices}}\sim N^{\sharp\text{white vertices}}$$ while, by (\ref{order}), the contribution of the product of cumulants is of the order $$O(N^{\sharp\text{black vertices}-\sharp\text{edges}})$$ It follows that the contribution coming from trees, for which 
 $$\sharp\text{white vertices}+\sharp\text{black vertices}-\sharp\text{edges}=1$$   is of the order $O(N)$ while the contribution of other graphs is of lower order in $N$.
For $H$ a tree, the combinatorial factor is $\mu(H^\bullet)=\prod_i (-1)^{k_i-1}(k_i-1)!$ the product being over white vertices and $k_i$ being the number of black neighbours
 of the white vertex indexed by $i$. This follows from the computation of the chromatic polynomial for the complete graph and the formula for the joining of two graphs given by (\ref{join}).
In this case we can thus rewrite 
$$\frac{\mu(H^\bullet)}{|\text{Aut}\,H|}w({\mathcal L})=\frac{1}{|\text{Aut}(H)|}\prod_\bullet K_\bullet(b_\bullet)\prod_{\circ}(-1)^{k_\circ -1}(k_\circ -1)!\prod_ee_i$$
where the product $\prod_\circ$ is over white vertices and $k_\circ$ denotes the number of neighbours of the white vertex $\circ$.

\subsubsection{Gradient of the free energy}\label{gq}
Let us now compute $e_i\frac{\partial W}{\partial e_i}$ in the large $N$ limit. Since, in the weight $w({\mathcal L})$, there is a factor $e_i$ for each edge adjacent to a white vertex labelled $i$,  one sees that this derivative is given by a sum over pairs $T,e$ of a tree $T$ and an edge $e$ of $T$

\begin{equation}\label{deW}e_i\frac{\partial W}{\partial e_i}
\sim\sum_T\frac{\mu(T^\bullet)}{|\text{Aut}\,T|}\sum_{e\,\text{edge of}\, T}\sum_{{\mathcal L};e_\circ\sim i}w({\mathcal L})
\end{equation}
where we sum over all labellings $\mathcal L$ such that the white vertex of $e$ is labelled by $i$. Cutting the edge $e$ splits the tree into two rooted trees,  $T_\bullet$ and $T_\circ$,  one of them containing the black vertex of $e$ as a root and  the other the white vertex, with labellings ${\mathcal L}_\bullet,{\mathcal L}_\circ$. The automorphism subgroup fixing $e$ is the product of the automorphism groups of these two rooted trees (that is, the automorphisms fixing the roots):
 $\text{Aut}_e(T)\sim{\text{Aut}(T_\bullet})\times \text{Aut}(T_\circ)$, while the term $$\prod_\bullet K_\bullet(b_\bullet)\prod_{\circ}(-1)^{k_\circ -1}(k_\circ -1)!e_i\prod_{e'\ne e}e_j$$ splits into a product over the two trees.
One can sum over all edges in the  orbit of $e$  by $\text{Aut}(T)$ (whose size is $|\text{Aut}(T)|/|\text{Aut}_e(T))$ and get a sum 
\begin{equation}\label{deW2}e_i\frac{\partial W}{\partial e_i}
\sim\sum_T\sum_{e\in E(T)/\text{Aut}(T)}\frac{1}{|\text{Aut}(T_\bullet)||\text{Aut}(T_\circ)|}\sum_{{\mathcal L};
 e_\circ\sim i}z_i^\bullet({\mathcal L}_\bullet)z_i^\circ({\mathcal L}_\circ)
\end{equation}
In this sum the weights $z_i^\bullet,z_i^\circ$ are computed on labelled bipartite trees with a black (resp. white) root and one has
 $$z_i^\bullet({\mathcal L}_\bullet)=e_iK_{root}(b_i,\ldots)\prod_{\bullet\ne root} K_\bullet(b_\bullet)\prod_{\circ}(-1)^{k_\circ -1}(k_\circ -1)!\prod_{e'}e_j$$
 for a tree with a black root, where  $K_{root}(b_i,\ldots)$ is the non-coincident cumulants evaluated on $b_i$ and the neighbours of the black root. Similarly 
 $$z_i^\circ({\mathcal L}_\circ)=(-1)^{k_{root}}k_{root}!\prod_{\bullet} K_\bullet(b_\bullet)\prod_{\circ\ne root}(-1)^{k_\circ -1}(k_\circ -1)!\prod_{e'}e_j$$
 for a tree with a white root.
 Let us introduce the functions 
 \begin{equation}\label{defq}
 q_i=\sum_{T_{\circ}}\frac{1}{|\text{Aut}(T_\circ)|}\sum_{{\mathcal L}_\circ}z_i^\circ({\mathcal L}_\circ)
 \end{equation}
 \begin{equation}\label{defg}
 g_i=\sum_{T_{\bullet}}\frac{1}{|\text{Aut}(T_\bullet)|}\sum_{{\mathcal L}_\bullet} z_i^\bullet({\mathcal L}_\bullet)
 \end{equation}
If we compare the expression on the rhs of (\ref{deW2}) with the product $g_iq_i$ we see that they coincide up to possibly some repetitions in the 
labellings in the expansion of $qg$ (since we consider the product of labellings of $T_\bullet$ and $T_\circ$), however the number of terms with repetition in the labellings is of smaller order in $N$
therefore, as $N\to\infty$, one has 

$$e_i\frac{\partial W}{\partial e_i}\sim g_i q_i.$$ 

One can depict the trees and the weights involved in the definition of the functions $g$ and $q$ as follows: 
$$
\begin{tikzpicture}[scale=1]
\draw[fill] (0,0) circle (.4cm);
\draw[color=red] (0.28,.28)--(2,2);\draw[color=red] (.4,0)--(3,0);\draw[color=red] (.28,-.28)--(2,-2);
\node[color=blue] at (2.3,2.1){$\ldots$};\node[color=blue] at (3.3,0){$\ldots$};\node[color=blue] at (2.3,-2.1){$\ldots$};
\node at (-2,0){$g_i$};
\node at (-2,-2){$T_\bullet$};
\node at (-.7,.6){$\scriptstyle\bf e_iK_k(b_i,\ldots,\ldots)$};
\node[color=red] at (1.2,1.6){$\scriptstyle e_{\ldots}$};\node[color=red] at (1.5,.2){$\scriptstyle e_{\ldots}$};
\node[color=red] at (1.3,-1.7){$\scriptstyle e_{\ldots}$};
\draw(4,-3)--(4,3);
\draw[color=blue] (6,0) circle (.4cm);
\node[color=blue] at (6,0){$i$};
\draw[color=red] (6.28,.28)--(8,2);\draw[color=red] (6.4,0)--(9,0);\draw[color=red] (6.28,-.28)--(8,-2);
\node at (8.3,2.1){$\bf\ldots$};\node at (9.3,0){$\bf \ldots$};\node at (8.3,-2.1){$\bf \ldots$};
\node[color=red] at (7.2,1.6){$\scriptstyle e_i$};\node[color=red] at (7.5,.2){$\scriptstyle e_i$};
\node[color=red] at (7.3,-1.7){$\scriptstyle e_i$};\node[color=red] at (7.3,-1.7){$\scriptstyle e_i$};
\node at (4.6,0){$q_i$};
\node at (4.6,-2){$T_\circ$};
\node[color=blue]  at (5.3,.6){$\scriptstyle\bf (-1)^{k_\circ}k_\circ !$};
\end{tikzpicture}
$$

\subsubsection{A variational principle}
Reasoning as in (\ref{gq}) by cutting the tree in either (\ref{defq}) or (\ref{defg}) at its root to form a forest, we find the following 
relations in the continuous limit between $q(s),g(s),e(s):=e^{h(s)}-1$:
$$
\begin{array}{rcl}
e(s)\frac{\partial W}{\partial e(s)}&=&g(s)q(s)\\
q(s)&=&\frac{e(s)}{1+e(s)g(s)}\\
g(s)&=&\frac{\delta}{\delta q(s)}F_0(q)
\end{array}
$$
where $F_0(q)$ is the large $N$ limit of the generating functional of the cumulants:

$$F_0(q)=\sum_n\frac{1}{n!}\int_{\Sigma^n} q(s_1)q(s_2)\ldots q(s_n)\psi_n(s_1,\ldots,s_n)ds_1\ldots ds_n$$
It follows that

\begin{prop}
In the scaling limit, the free energy is obtained by solving the following  variational problem
\begin{equation}
\lim_{N\to\infty}\frac{1}{N}W=\max_{g,q}\left[\int [\log(1+e(s)g(s))-q(s)g(s)] ds+F_0(q)\right]
\end{equation}
\end{prop}
We shall use this variational formula in the case of SSEP in Section \ref{sec:SSEP}.

\section{Bernoulli Partition Functions and Feynman Graphs}
\label{sec:Feynman}

In this section we employ standard field theory techniques (mainly the semi-classical expansion and Feynman graphs) to encode the combinatorics of Bernoulli cumulants. 

\subsection{Integral representation of $\log Z$}\label{ssec:FeynmanZ}


We start with another description of $\log Z$ in terms of a graphical expansion. The starting point is a somehow tautological representation of $Z$ as a formal Gaussian integral (see \autoref{appssec:FGI} and the following for some background if needed). 

The basic observation is that, $J$ being an arbitrary index set, $({\overline \lambda}_i)_{i\in J}$ and $(\lambda_I)_{I \stackrel{\circ}{\subset}  J}$ being formal variables (the notation $I \stackrel{\circ}{\subset} J$ means that $I$ is a finite, nonempty subset of $J$):
\[ \int\left(\prod_{i\in J} \frac{d{\overline z}_i\wedge dz_i}{2i\pi} \exp (- z_i {\overline z}_i) (1+{\overline \lambda}_iz_i)\right) \exp (\sum_{I \stackrel{\circ}{\subset}  J} \lambda_I {\overline z}_I) = 1+\sum_{I \stackrel{\circ}{\subset}  J} {\overline \lambda}_I \left(\sum_{\pi \in {\mathcal P}(I)} \lambda_{\pi}\right),\]
where ${\overline \lambda}_I:=\prod_{i\in I} {\overline \lambda}_i$ (and analogously ${\overline z}_I:=\prod_{i\in I} {\overline z}_i$), ${\mathcal P}(I)$ is the set of partitions of $I$ and, for $I\neq \emptyset$ and $\pi:I=\sqcup_{\alpha}  I_{\alpha} \in {\mathcal P}(I)$, $\lambda_{\pi}:= \prod_{\alpha} \lambda_{I_{\alpha}}$.
This formula is checked by expanding
\[ \prod_{i\in J} (1+{\overline \lambda}_iz_i)=1+\sum_{I \stackrel{\circ}{\subset}  J} {\overline \lambda}_Iz_I.\]
Concentrating on a given ${\overline \lambda}_I$, formal integration amounts to selecting, in the expansion of $\exp (\sum_{I' \stackrel{\circ}{\subset}  J} \lambda_{I'} {\overline z}_{I'})$, precisely the terms involving the monomial ${\overline z}_I$, which by inspection come with an overall factor $\sum_{\pi \in {\mathcal P}(I)} \lambda_{\pi}$.

Though this is a formal integral, if the index set $J$ is finite the result is a polynomial in $({\overline \lambda}_i)_{i\in J}$ and $(\lambda_I)_{I \stackrel{\circ}{\subset}  J}$, and can be evaluated for ``numerical'' arguments. As established in the previous section, if $b_i$ are Bernoulli variables then
\[ Z={\mathbb E}(\prod_{i\in J} (1+e_ib_i))=1+\sum_{I \stackrel{\circ}{\subset}  J} e_i \sum_{\pi \in {\mathcal P}(I)} K_{\pi}(b_I),\]
with notations as above, 
Thus
we may write
\[ Z= \int\left(\prod_{i\in J} \frac{d{\overline z}_i\wedge dz_i}{2i\pi} \exp (- z_i {\overline z}_i) (1+e_iz_i)\right) \exp (\sum_{I \stackrel{\circ}{\subset}  J} K(b_I) {\overline z}_I).\]
Thus in the sequel we do not distinguish between the formal variables $({\overline \lambda}_i)_{i\in J}$, $(\lambda_I)_{I \stackrel{\circ}{\subset}  J}$ and their embodied counterparts $(e_i)_{i\in J}$, $(K(b_I))_{I \stackrel{\circ}{\subset}  J}$, which we shall use in the formul\ae . 

We can turn the crank of Feynman graphs and rules as recalled in \autoref{appsec:FGR} and \autoref{appssec:SC}. Writing
\[ \prod_{i\in J} (1+e_iz_i)=\exp \sum_{i\in J}\sum_{k\geq 1} \frac{(-1)^{k-1}(k-1)!e_i^kz_i^k}{k!} \]
we infer that
\[ \log Z = \sum_G w(G) \]
where the sum is over connected bicolored graphs with white and black vertices whose edges carry a type $i\in J$, the type of the edges at a white vertex being all the same $(\circ)$ and at a black vertex all different $(\bullet)$. Each white vertex with $k$ edges of type $i\in J$ contributes a factor $(-1)^{k-1}(k-1)!e_i^k$ to $w(G)$. Each black vertex with edges whose types build a subset $I \stackrel{\circ}{\subset} J$ contributes a factor $K_I$ to $w(G)$. There a an additional factor $1/ |\text{Aut}\, G|$ in $w(G)$. The two constraints $(\circ)$ and $(\bullet)$ allow to ``transfer'' the types of edges to the white vertices and to consider connected bicolored graphs with white and black vertices, the white vertices carrying a tag $i\in J$ and the tags of white vertices connected to a black vertex being different.

We give some examples, assuming that $J$ is a set of integers:\\
-- Examples: A white vertex of order $4$ associated to site $i=3$, with weight $(-1)^{4-1}(4-1)!e_3^4=-6e_3^4$, a white vertex of order $3$ associated to site $7$, with weight $(-1)^{3-1}(3-1)!e_7^3=2e_7^3$ and a black vertex connected to sites $1,3,9,5$ with weight $K_4(b_1,b_3,b_9,b_5)$:
\begin{center}
\includegraphics[width=.6\linewidth]{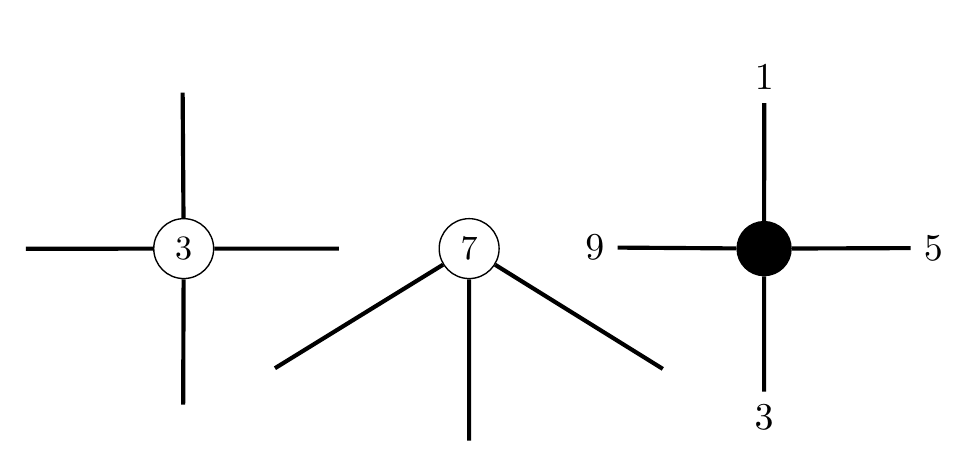}
\end{center}
-- Example: A diagram 
\begin{center}
\includegraphics[width=.5\linewidth]{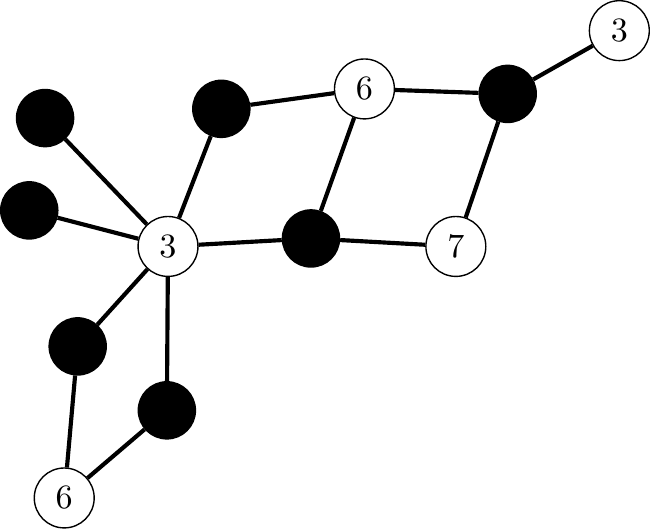}
\end{center}
with weight (reading more or less from left to right, in that simple case symmetries are ``local'' on the graph)
\[\frac{1}{2!}K_{3}^2 (-1)^{2-1}(2-1)!e_6^2 \frac{1}{2!} K_{3,6}^2 
   (-1)^{6-1}(6-1)!e_3^6 K_{3,6}K_{3,6,7}\] \[ (-1)^{3-1}(3-1)!e_6^3 (-1)^{2-1}(2-1)!e_7^2 K_{3,6,7}(-1)^{1-1}(1-1)!e_3,\]
where $K_{3}$, $K_{3,6}$, $K_{3,6,7}$ are a shorthand notation for the cumulants $K_1(b_3)$, $K_2(b_3,b_6)$ and $K_3(b_3,b_6,b_7)$ respectively, a convention already used above.

 The weight reduces to
$-60 e_3^7e_6^5e_7^2K_3^2K_{3,6}^3K_{3,6,7}^2$.

From now on, we could reproduce with very little changes the discussion leading to the continuum limit, to which only tree diagrams contribute.\\
-- Here is an example of a tree: 
\begin{center}
\includegraphics[width=.5\linewidth]{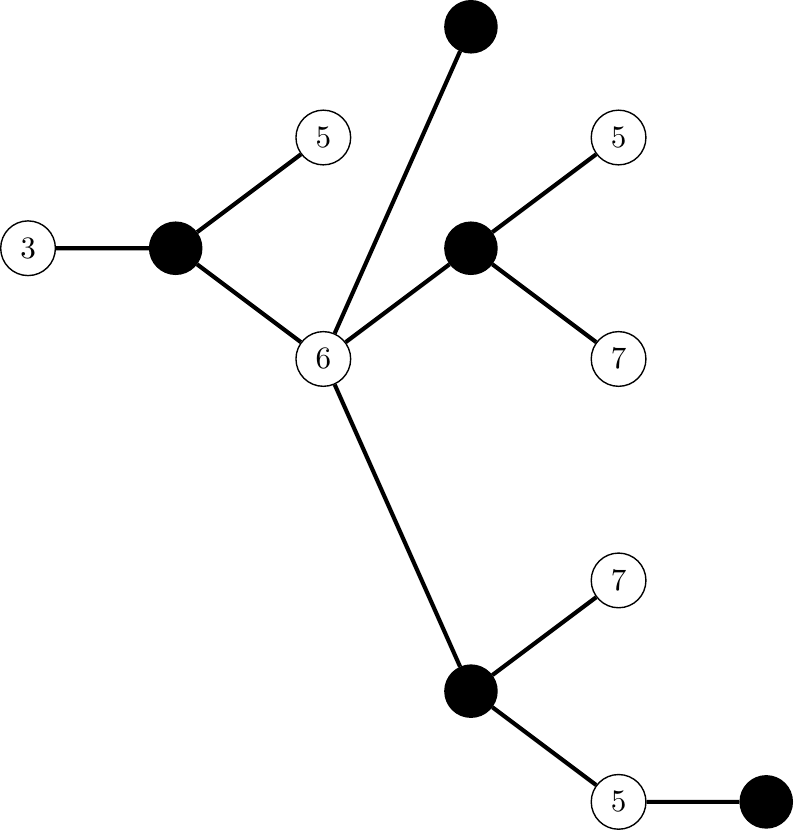}
\end{center}
the computation of whose weight is left to the reader.

We pause for a moment to compare with the other (call it \textit{chromatic}) graphical description of $\log Z$ given in \autoref{ssec:CBv}.

For this, we need to rewrite some previous formulas, in particular eq. (\ref{eq:weight-L0}) and (\ref{eq:weight-L}). In the chromatic description, instead of summing over unlabelled graphs $H$ and then over labelings by (distinct) elements of $J:=\{1,\cdots,N\}$, we can as well sum over graphs $G$ whose white vertices are labelled by $\{1,\cdots,N\}$. If $G$ is such a graph, define ${\mathfrak w}(G):=\prod_{\bullet}K(b_\bullet)\prod_{\text{edges of $G$}}e_i$,
so that, if $G$ is obtained from an unlabelled graph $H$ via the labelling ${\mathcal L}$, ${\mathfrak w}(G)=w({\mathcal L})$. A moment thinking shows that 
\[ \frac{1}{|\text{Aut}\, H|} \sum_{{\mathcal L}\in Lab(H)}w({\mathcal L})= \sum_{G} \frac{1}{|\text{Aut}\, G|} {\mathfrak w}(G) \]
where the sum on the right-hand side is over the distinct graphs $G$ that can be obtained by labeling the white vertices of $H$, and $\text{Aut}\, G$ is the group of automorphisms of $G$ respecting the labeling, which may be smaller than $\text{Aut}\, H$ because several labelings of $H$ may induce the same $G$, as for instance in the example (\ref{eq:graphil}). As all white labels of $G$ are distinct, the automorphism group of $G$ is in fact very easy to describe: saying that two black vertices are equivalent if they are connected to the same set of white vertices, $\text{Aut}\, G$ is the group of permutations of equivalent black vertices. The definition of the operation $^\bullet$ for $G$ can be copied on that of $H^\bullet$ and clearly $G^\bullet=H^\bullet$. Thus the complete contribution of $G$ to the free energy $\log Z$ in the chromatic description is $\frac{\mu (G^\bullet)}{|\text{Aut}\, G|} {\mathfrak w}(G)$. 

The graphs $G$ we have used for the Feynman graph description have a tag, an element of $J=\{1,\cdots,N\}$, assigned to each white vertex, but this is not a labeling of the white vertices in general because several white vertices with the same tag are allowed.\footnote{However, the tags of white vertices adjacent to a given black vertex are different.} Consequently, their automorphism group is more complicated to describe in general because it may permute white vertices as well. But the definition of the monomial ${\mathfrak w}(G)$ carries over without changes for Feynman graphs. On top of that, a white vertex of order $k$ contributes a multiplicative factor $(-1)^{k-1}(k-1)!$. Defining $G^\circ$ as the collection of white vertices of $G$ with their pending edges (i.e. one just removes the black vertices from the picture) and $\eta(G^\circ):= \prod_{\text{white vertices of } G}(-1)^{k-1}(k-1)!$, we can put things together and write the complete contribution of $G$ to the free energy $\log Z$ in the Feynman graph description as $\frac{\eta (G^\circ )}{|\text{Aut}\, G|} {\mathfrak w}(G)$.

Working as above with (connected) graphs whose white vertices are tagged, let us denote by  ${\mathcal C}_C$ (resp. ${\mathcal C}_F$) the class of graphs involved in the chromatic (resp. Feynman graph) expansion of $\log Z$. We have just seen that ${\mathcal C}_C\subset {\mathcal C}_F$: the chromatic graphical description which is tailored for the problem at hand and is more economical because there are less graphs to consider. We have
\[ \log Z = \sum_{G \in {\mathcal C}_C} \frac{\mu (G^\bullet)}{|\text{Aut}\, G|} {\mathfrak w}(G)=\sum_{G \in {\mathcal C}_F} \frac{\eta (G^\circ )}{|\text{Aut}\, G|} {\mathfrak w}(G). \]

We may refine this identity using the obvious observation that for any finite collection (possibly with repetition) of non-empty finite subsets of $J$, say $(I_a)_{a\in A}$ there is a single graph $G$ in ${\mathcal C}_C$ with black vertices indexed by the $I_a$ i.e. ``black'' weight $\prod_{a\in A} K_{I_a}$. In particular a graph $G\in {\mathcal C}_C$ can be reconstructed from the sole knowledge of ${\mathfrak w}(G)$, leading to the identity 
\[ \frac{\mu (G^\bullet)}{|\text{Aut}\, G|}=\sum_{G'\in {\mathcal C}_F,\, {\mathfrak w}(G')={\mathfrak w}(G)}  \frac{\eta ((G')^\circ )}{|\text{Aut}\, G'|}\quad \text{for } G\in {\mathcal C}_C.\]

This result is perhaps more suggestive if one introduces a partial ordering on ${\mathcal C}_F$: for $G,G' \in {\mathcal C}_F$ say that $G'\succcurlyeq G$, or that $G'$ covers $G$ if $G$ is obtained from $G'$ by identifying some white vertices carrying the same tag. This is clearly a partial ordering on ${\mathcal C}_F$. The maximal elements are the trees and the minimal elements are the elements of ${\mathcal C}_C$. If $G'\succ  G$, $G$ has less white vertices than $G'$ and $G$ has more loops than $G'$. The graphs $G$ and $G'$ have the same number of black vertices and the same number of edges, and in fact the equality ${\mathfrak w}(G)={\mathfrak w}(G)$ holds. In particular, given $G$ there are finitely many $G'\succcurlyeq G$ and the previous identity rewrites
\begin{equation} \label{eq:identCF}
  \frac{\mu (G^\bullet)}{|\text{Aut}\, G|}=\sum_{G'\in {\mathcal C}_F, G' \succcurlyeq G} \frac{\eta ((G')^\circ )}{|\text{Aut}\, G'|}\quad \text{for } G\in {\mathcal C}_C.\end{equation}

The above argument gives a rigorous but indirect proof of this identity. A direct proof for trees is easy: if $G$ in ${\mathcal C}_C$ is a tree, there is only $G$ itself in the sum on the right-hand side ($G$ is minimal and maximal for $\succcurlyeq$),\footnote{Thus, though the number of terms on the right-hand side of (\ref{eq:identCF}) can be arbitrarily large for a general $G\in {\mathcal C}_C$, the overhead of using the Feynman graph description disappears in the thermodynamic/continuous limit.} and the  $(-1)^{k-1}(k-1)!$s coming from complete graphs chromatic factors in $\mu (G^\bullet)$ match precisely with Feynman graph contributions for white vertices in $\eta ((G)^\circ )$. Note however that the ``Feynman trees'' allow for several white vertices with the same tag. They cover graphs with loops in the chromatic expansion. In both the chromatic and the Feynman graph expansion, loops are suppressed with respect to trees when the number of sites, $|J|$, grows without bounds. This is not in contradiction with the semi-classical expansion: for a fixed number of white vertices, the factor suppressing trees with multiple vertices carrying the same tag is just due to their rarity compared to trees with all white vertices carrying a different tag, and this matches precisely with the scaling in the covering formula. The partial order $\succcurlyeq$ on ${\mathcal C}_F$ suggest that a recursive approach might help to give a direct proof of  (\ref{eq:identCF}) in general, but we have not tried to follow this path.

A final remark: the computation of $|\text{Aut}\, G|$ in the class ${\mathcal C}_F$ is $NP$-hard, just as is the computation of $\mu (G^\bullet)$ in the class ${\mathcal C}_C$, whereas the computation of $\eta (G^\circ)$ in ${\mathcal C}_F$ or $|\text{Aut}\, G|$ in ${\mathcal C}_C$ is trivial. 

\subsection{Thermodynamic limits}\label{ssec:TL}

One small thing that speaks in favor for the redundant description of $\log Z$ by Feynman graphs is that it generalizes plainly to related counting problems.

Suppose for instance that $J$ is finite, that $e_i$ does not depend on $i$ and $K_I$ depends only on the size $|I|$ of $I\subset J$ (without bothering if this can happen for actual Bernoulli variables expectations, in fact it does at least for the trivial case of independent identically distributed Bernoulli when $e_i$ does not depend on $i$ because  $K_I$ 
is the $|I|$th power of a single variable expectation). We fix a family $(t_k)_{k\geq 1}$ of formal variables and set $K_I=:t_k$ if $|I| =k$. For $I \stackrel{\circ}{\subset}  J$ the number of partitions of $I$ made of $m_1$ parts of size $1$, $m_2$ parts of one size $2$, and so on with $\sum_{k\geq 1} km_k =|I|$ is $\frac{(\sum_{k\geq 1} km_k)!}{\prod_{k \geq 1} m_k! (k!)^{m_k}}$, leading to 
\[ Z=\sum_{\underline m}  \frac{|J| !}{(|J|-\sum_{k\geq 1} km_k)!} \prod_{k\geq 1} \frac{1}{m_k!}\left(\frac{e^kt_k}{k!}\right)^{m_k} \]
where $e_i:=e$ and $K_I=:t_{|I|}$ and the sum is over sequences of integers $\underline m:=(m_1,m_2,\cdots)$. The point is that the combinatorics is precisely recovered in a formal Gaussian integral as
\[ Z = \int \frac{d{\overline z}\wedge dz}{2i\pi\hbar} \exp \left(- \frac{z{\overline z}}{\hbar}\right)\, (1+ez)^{|J|} \exp \sum_{1 \leq k \leq |J|} \frac{t_k}{k!} \left(\frac{\overline z}{\hbar}\right)^k. \]
Notice that $\hbar$, whether a formal variable or a numerical value, plays a purely spectator role in this formula. The expansion of $Z$ in terms of Feynman graphs would follow straightforwardly. 

We use rescaled variables $\Phi_k=:(|J|)^{k-1} t_k$ for $k \geq 1$. Choosing $\hbar:=(|J|)^{-1}$ and precising the variables involved in $Z$ we are led to
\[ Z_{|J|}(e,\Phi_\mybullet) = \int \frac{d{\overline z}\wedge dz}{2i\pi (|J|)^{-1}} \exp \frac{1}{(|J|)^{-1}}\left( - z{\overline z}+\log (1+ez) + \sum_{k \geq 1 } \frac{\Phi_k}{k!} {\overline z}^k\right).\] 
Thus $(|J|)^{-1}$ plays the role that $\hbar$ plays in the general discussion of \ref {appssec:SCE} and \ref{appsec:FGR}  and letting $|J|$ grow without bounds with $\Phi_\mybullet:=(\Phi_k)_{k\geq 1}$ fixed we obtain that in the thermodynamic limit 
\[ \lim_{|J| \to \infty} \frac{\log Z_{|J|}(e,\Phi_\mybullet)}{|J|} =F^*(e,\Phi_\mybullet)   \]
where $F^*(e,\Phi_\mybullet)=-g^*h^*+\log(1+eg^*) +\sum_{k\geq 1} \Phi_k \frac{{h^*}^k}{k!}$ with $(g^*,h^*)$ solving the equations $h=\frac{e}{1+eg}$ and $g=\sum_{k \geq 1 } \Phi_k \frac{h^{k-1}}{(k-1)!}$. The formal power series $g$ (resp. $h$) is what we denoted by $U$ (resp. ${\overline V}$) in the general discussion of \autoref{appssec:SCE} and \autoref{appssec:SC}.

As a slight generalization of this extreme case, suppose that $J$ is finite and $J=\cup_{a\in A} J_a$ is a partition of $J$ indexed by some set $A$. Impose that $e_i$ is the same for all $i$s in a given $J_a$, and write it as $e_a$. Fix a family $(t_k)_{k\geq 1}$ where each $t_k$ is a symmetric function on $A^k$ and for $I \stackrel{\circ}{\subset}  J$ set $K_I=t_k(a_1,\cdots,a_k)$  if $I=\{i_1,\cdots,i_k\}$ and $i_l\in J_{a_l}$ for $l=1,\cdots,k$. The extreme case is recovered when $A$ is a singleton. The counting of partitions in the extreme case generalizes straightforwardly and leads to the integral representation
\[ Z = \int  \left(\prod_{a\in A} \frac{d{\overline z}_a\wedge dz_a}{2i\pi\hbar} \exp (- z_a {\overline z}_a/\hbar) (1+e_az_a)^{|J_a|} \right) \exp {\overline {\mathcal L}}_{|J|}({\overline z}_\mybullet)\]
where
\[ {\overline {\mathcal L}}_N({\overline z}_\mybullet):= \sum_{1 \leq k \leq N} \frac{1}{\hbar^k k!}\sum_{a_1,\cdots,a_k\in A} t_k(a_1,\cdots,a_k){\overline z}_{a_1}\cdots{\overline z}_{a_k}.\] 
Again $\hbar$ is a spectator role in this representation. We use rescaled variables $\Phi_k=:(|J|)^{k-1} t_k$ for $k \geq 1$. Choosing $\hbar:=(|J|)^{-1}$ and precising the variables involved in $Z$ we are led to
\[ Z_{|J|}(e_\mybullet,\Phi_\mybullet)=\int \prod_{a\in A} \frac{d{\overline z}_a\wedge dz_a}{2i\pi(|J|)^{-1}}  \exp \frac{1}{(|J|)^{-1}}\left(\sum_{a\in A} (- z_a {\overline z}_a + p_a \log (1+e_az_a)) +{\overline L}_{|J|} ({\overline z}_\mybullet)\right),\]
where $p_a:=\frac{|J_a|}{|J|}$ for $a\in A$ and
\[ {\overline L}_N({\overline z}_\mybullet):= \sum_{1 \leq k \leq N} \frac{1}{k!}\sum_{a_1,\cdots,a_k\in A} \Phi_k(a_1,\cdots,a_k){\overline z}_{a_1}\cdots{\overline z}_{a_k}.\]
Thus, there is a semi-classical expansion in powers of $(|J|)^{-1}$, with $A$, $\Phi_\mybullet$ and $(p_a)_{a\in A}$ fixed. The first contribution, proportional to $|J|$ is given by the saddle point $F^*(e_\mybullet,\Phi_\mybullet)$ where $F^*(e_\mybullet,\Phi_\mybullet)=-\sum_{a\in A} g_a^*h_a^*+\sum_{a\in A} p_a\log (1+e_ag_a^*) +{\overline L}_\infty (h^*_\mybullet)$ with $(g_a^*,h_a^*)_{a\in A}$ solving the equations $h_a=\frac{e_a}{1+e_ag_a}$ and $g_a=\frac{\partial {\overline L}}{\partial {\overline z}_a}(h_\mybullet)$, $a\in A$. Thus, letting $|J|$ grow without bounds with $A$, $\Phi_\mybullet$ and $(p_a)_{a\in A}$ fixed (this might require taking $|J|$ along some subsequence) we obtain
\[ \lim_{|J| \to \infty} \frac{\log  Z_{|J|}(e_\mybullet,\Phi_\mybullet)} {|J|} =F^*(e_\mybullet,\Phi_\mybullet)\]
The full semi-classical expansion is valid only for $(p_a)_{a\in A}$ fixed, but this limiting result holds if $(p_a)_{a\in A}$ depends on $|J|$ with corrections $o((|J|)^{-1})$ and $|J| \to \infty$ without restrictions.

To make contact with the general formul\ae , take  $(f_k)_{k\geq 1}$ a sequence of symmetric integrable functions, $f_k:[0,1]^k\to {\mathbb R}$ and consider the functional
\begin{eqnarray*} F(e,f_\mybullet,g,q)& := &-\int_{[0,1]} g(x)\, q(x)dx+\int_{[0,1]} \log(1+e(x)g(x))\, dx \\ & & +\sum_{k\geq 1}\int_{0<x_1<\cdots <x_k} f_k(x_1,\cdots,x_k) \, q(x_1)dx_1\cdots \, q(x_k)dx_k ,\end{eqnarray*}
where $e,g,q$ are plain functions. Note that the multiple integral could also be written as $\frac{1}{k!} \int_{[0,1]^k}f_k(x_1,\cdots,x_k) \, q(x_1)dx_1\cdots \, q(x_k)dx_k$ and the diagonals (where several $x$'s coincide) do not contribute.
The general formul\ae\ come via the extremization of $F$ with respect to the functions $g$ and $q$. One way to approximate this problem is to partition $[0,1]$ with measurable subsets $(M_a)_{a\in A}$ with size $\int_{M_a}\, dx =p_a>0$ for $a\in A$ and take $e,g,q$ as simple functions, explicitly 
\[e(x)=\sum_a e_a \ind{x\in M_a} \quad g(x)=\sum_a g_a \ind{x\in M_a} \quad q(x)=\sum_a q_ap^{-1}_a \ind{x\in M_a}.\]
Taking $\Phi_\mybullet$ as the average of $f_\mybullet$ over rectangles, explicitly
\[ \Phi_k(a_1,\cdots,a_k):=\frac{1}{p_{a_1}\cdots p_{a_k}} \int_{M_{a_1}\times \cdots \times M_{a_k}} f_k(x_1,\cdots,x_k) \,dx_1\cdots \, dx_k,\]
(the diagonals where several $a_l$s may coincide do count in the discrete object $\Phi_\mybullet$ ) we find that the discretized version of $F(e,f_\mybullet,g,q)$ is
\[
 -\sum_{a\in A} g_aq_a+\sum_{a\in A} p_a\log (1+e_ag_a)+\sum_{k \geq 1} \frac{1}{k!}\sum_{a_1,\cdots,a_k\in A} \Phi_k(a_1,\cdots,a_k)q_{a_1}\cdots q_{a_k} ,\]
which coincides with the thermodynamic limit functional above, to be extremized with respect to $(g_a,q_a)_{a\in A}$. This paves the way to another approach to the general formul\ae , via the approximation of the $K(b_I)$, $I \stackrel{\circ}{\subset} J$ by step functions.

\section{Classical SSEP and Free Probability}
\label{sec:SSEP}

\subsection{The classical SSEP}

The aim of this section is to recall the definition of the classical SSEP and its large deviation function. Results are taken from the SSEP literature \cite{Derrida_Review,Mallick_Review} where further details may be found.

The classical SSEP is a time continuous Markov chain describing particles moving along a finite 1D lattice, with sites indexed by $i=1,\cdots, N$ (the sites $i$ and $i+1$ are adjacent). Let $\tau_i$ be the occupation number of the site $i$~: $\tau_i=0$ (resp. $\tau_i=1)$ if the site $i$ is unoccupied (resp. occupied). Each configuration is specified by the data of these occupancies $\{\tau_i;\ i=1,\cdots,N\}$. The particles are allowed to jump on their nearby positions, to their left or right with equal probability rate, if the target position is unoccupied. The allowed local moves are therefore $[01]\to [10]$ or $[10]\to [01]$, while the local configurations $[00]$ and $[11]$ are frozen. Particles are injected and extracted at the two ends of the interval to drive the system out-of-equilibrium. The SSEP Markov matrix is defined accordingly to take these moves into account in a natural way. We denote by $\mathbb{E}_\mathrm{ssep}$ the SSEP invariant measure (which is known to be unique).

One is interested in the continuum scaling limit $N\to\infty$, $x=i/N$ fixed, $0<x<1$. The occupation configurations $\{\tau_i\}$ then become continuous density profiles $\mathfrak{n}(x)$ on the interval $[0,1]$. In this scaling limit, the densities at the two ends of the interval are fixed by the injection-extraction processes~: $\mathfrak{n}(0)={n}_a$ and $\mathfrak{n}(1)={n}_b$ with $n_a$ and $n_b$ specified by the injection-extraction rates at the corresponding boundary. We shall use the convention $n_a=0$, $n_b=1$ (without loss of generality).

In the scaling limit, the mean density profile interpolates linearly between the two boundary densities~: $\lim_{N\to\infty}\mathbb{E}_\mathrm{ssep}[\tau_{i=[xN]}]=x$ (with the convention $n_a=0$, $n_b=1$). Fluctuations of the density profiles satisfy a large deviation principle~\cite{DerridaLarge}. Namely, 
\beq \label{eq:Issep-def}
\mathbb{P}\left[\tau_{i=[xN]}\approx n(x)\right] \asymp_{N\to \infty} e^{-N \,I_\mathrm{ssep}[n] } 
\eeq
with $I_\mathrm{ssep}$ the so-called large deviation rate function.

Let $F_\mathrm{ssep}$ be the generating function of the density cumulants in the scaling limit. It is such that 
\beqa \label{eq:Wh-dev}
\mathbb{E}_{\mathrm{ssep}}\big[ e^{ \sum_i \tau_i h_i } \big] \asymp_{N\to \infty} e^{N\,F_\mathrm{ssep}[h] } ~,
\eeqa
for $h_i=h(x=i/N)$ with $h(x)$ a smooth function over the interval $[0,1]$. If the occupation configuration $\{\tau_i\}$ approaches the density profiles $\mathfrak{n}(x)$, then $\sum_i \tau_i h_i$ approches $N \int \! dx\, h(x)\mathfrak{n}(x)$, so that $F_\mathrm{ssep}$ can alternatively be defined by $\mathbb{E}_{\mathrm{ssep}}\big[ e^{ N \int \! dx\, h(x)\mathfrak{n}(x) } \big] \asymp_{\varepsilon\to 0} e^{N F_\mathrm{ssep}[h] }$. Assuming eq.\eqref{eq:Wh-dev} to be true implies that that the $p$-th order density cumulants scale like $N^{1-p}$ in the scaling limit. 

As is well known, the two functions $I_\mathrm{ssep}[n]$ and $F_\mathrm{ssep}[h]$ are related by Legendre transform~: 
\beq
F_\mathrm{ssep}[h] = \max_{n(\cdot)} \Big[ \int_0^1\!\!  dx\, h(x)n(x) - I_\mathrm{ssep}[n]\Big] .
\eeq

The large deviation function $F_\mathrm{ssep}[h]$ has been given in the SSEP literature \cite{DerridaLarge,Derrida_Review,Mallick_Review} as the solution of an extremization problem (with the convention $n_a=0$, $n_b=1$)~: 
\beqa \label{eq:Wh-int}
 && F_\mathrm{ssep}[h] = \max_{g(\cdot)}\, F[h;g], 
 \quad  F[h;g]:=  \int_0^1 \!\!\! dx\left[ \log\big(1+g(x)e(x)\big) - \log({g'(x)})\right] , 
 \eeqa
with $e(x)=e^{h(x)}-1$, as above, and $g(x)$ solution of the non-linear differential equation,
\beqa  \label{eq:Wh-extrem}
 \big(1+g(x)\,e(x))\big)\, g''(x) = g'(x)^2\, e(x) ~, 
 \eeqa
with boundary conditions $g(0)=0$ and $g(1)=1$ (for $n_a=0$, $n_b=1$). Eq.\eqref{eq:Wh-extrem} is the Euler-Lagrange equation for $F[h;g]$ to be extremal with respect to variations of $g$. 

Expanding $F_\mathrm{ssep}[h]$ in power of $h$ yields the first few density cumulants (up to sub-leading terms in $1/N$)~: 
\beqa \label{eq:corr-ssep}
&&\mathbb{E}_\mathrm{ssep}[\tau_{i=[xN]}\tau_{i=[yN]}]^c= - N^{-1}\, x(1-y)~,\nonumber\\
&& \mathbb{E}_\mathrm{ssep}[\tau_{i=[xN]}\tau_{i=[yN]}\tau_{i=[zN]}]^c= N^{-2}\, x(1-2y)(1-z), \nonumber
\eeqa
for $0<x<y<z<1$. More generally, the $n$-th order SSEP cumulants scale as $N^{1-n}$ in the scaling limit, see ref. \cite{Derrida_Review,Mallick_Review}. We let $\psi^\mathrm{ssep}_n(x_1,\cdots,x_n)$ be the scaled cumulants, at non-coincident points, defined as
\beq \label{eq:def-Cssep}
\psi^\mathrm{ssep}_n(x_1,\cdots,x_n) := \lim_{N\to\infty} N^{n-1}\mathbb{E}_\mathrm{ssep}\left[ \tau_{[x_1N]}\cdots \tau_{[x_nN]}\right]^c ,
\eeq
for $x_k\in[0,1]$, all distincts. The limit is known to exist and to be smooth (actually piecewise polynomial) at non-coincident points.


\subsection{The quantum to classical SSEP correspondance}

The aim of this section is to recall the definition of the quantum SSEP (Q-SSEP) as well as its relation with the classical SSEP. Results explained below are taken from ref.\cite{BernardJin19,BernardJin20} where further details may be found.

The quantum SSEP is a model of stochastic quantum dynamics describing fermions hopping along a 1D chain, with sites indexed by $i=1,\cdots, N$ (the sites $i$ and $i+1$ are adjacent). For an open chain in contact with external reservoirs at their boundaries, the quantum SSEP dynamics results from the interplay between unitary, but stochastic, bulk flows with dissipative, but deterministic, boundary couplings. The bulk flows induce unitary evolutions of the system density matrix $\rho_t$ onto $e^{-idH_t}\,\rho_t\,e^{idH_t}$ with Hamiltonian increments
\beq \label{eq:defHXXsto}
dH_{t}=\sqrt{J}\,\sum_{j=1}^{N}\big(c_{j+1}^{\dagger}c_{j}\,dW_{t}^{j}+c_{j}^{\dagger}c_{j+1}\,d\overline W_{t}^{j}\big),
\eeq
for a chain of length $L$, where $c_{j}$ and $c_{j}^{\dagger}$ are canonical fermionic operators, one pair for each site of the chain, with $\{c_{j},c_{k}^{\dagger}\}=\delta_{j;k}$, and $W_{t}^{j}$ and  $\overline W_{t}^{j}$ are pairs of complex conjugated Brownian motions, one pair for each edge along the chain, with quadratic variations $dW_{t}^{j}d\overline{W}_{t}^{k}=\delta^{j;k}\,dt$. The contacts with the external leads are modelled by Lindblad terms~\cite{Lindblad}. The resulting equation of motion for the system density matrix $\rho_t$ reads
\beq \label{eq:Q-flow}
 d\rho_{t}=-i[dH_{t},\rho_{t}] -\frac{1}{2}[dH_{t},[dH_{t},\rho_{t}]]+\mathcal{L}_\mathrm{bdry}(\rho_{t})dt,
 \eeq
with $dH_t$  as above and $\mathcal{L}_\mathrm{bdry}$ the boundary Lindbladian. The two first terms result from expanding the unitary increment $\rho_t \to e^{-idH_t}\,\rho_t\,e^{idH_t}$ to second order as indicated by It\^o calculus. The third term codes for the dissipative boundary dynamics representing injection-extraction at the two the boundaries. We do not need here the precise expression for $\mathcal{L}_\mathrm{bdry}$ but the latter can be found in the literature \cite{BernardJin19,Bernard21}.

The classical SSEP is embedded in Q-SSEP because the average Q-SSEP dynamics on density matrix diagonal in the particle number basis reduces to the classical SSEP. 
At each site along the chain, the full $\ket{\bullet}$ and empty $\ket{\oo}$ states, with respectively one and zero fermion, form a basis of states and diagonalize the particle number operators $\hat n_i:=c^\dag_ic_i$ (with eigen-value $1$ or $0$). The states $\ket{\bm{n}}$ diagonalizing all the particle numbers along the chain are thus indexed by the classical configurations $\bm{n}=(\tau_1,\cdots,\tau_N)$, with $\tau_i=0,1,$ the particle number at site $i$.
 A density matrix diagonal in this particle number basis specifies a probability measure on classical configurations since it can be written as $\rho_\mathrm{diag} = \sum_{\bm{n}} Q_{\bm{n}}\, \Pi_{\bm{n}}$, with $\Pi_{\bm{n}}:=\ket{\bm{n}}\bra{\bm{n}}$ the projector on the classical configuration $\bm{n}$ and $Q_{\bm{n}}$ a probability measure on $\bm{n}$~: $\sum_{\bm{n}} Q_{\bm{n}}=1$, $Q_{\bm{n}}\geq0$. 
 
 By the Markov property of the Brownian motions, the average dynamics deduced from eq.\eqref{eq:Q-flow} defines a semi-group on the average density matrix $\bar\rho_t:=\mathbb{E}[\rho_t]$, generated by a Lindbladian $\mathcal{L}_{\mathrm{ssep}}$~: 
\beq \label{eq:L1-dyn}
\partial_t \bar \rho_t = \mathcal{L}_{\mathrm{ssep}}(\bar \rho_t).
\eeq
The latter is obtained by averaging the Q-SSEP stochastic equation of motion \eqref{eq:Q-flow}. It preserves diagonal density matrices and thus defines a flow -- a Markov chain -- on probability measures on classical configurations. Locally $\mathcal{L}_{\mathrm{ssep}}$ acts as follows~:
\begin{align*}
\mathcal{L}_{\mathrm{ssep}} (\ket{\oo\oo}\bra{\oo\oo}) & =0 ~,\\
\mathcal{L}_{\mathrm{ssep}}(\ket{\oo\bullet}\bra{\oo\!\bullet\!}) & = J( -\ket{\oo\bullet}\bra{\oo\!\bullet\!}+\ket{\!\bullet\!\oo}\bra{\!\bullet\oo} )~,\\
\mathcal{L}_{\mathrm{ssep}}(\ket{\!\bullet\!\oo}\bra{\bullet\oo}) & = J( +\ket{\oo\bullet}\bra{\oo\!\bullet\!}-\ket{\!\bullet\!\oo}\bra{\!\bullet\oo} )~,\\
\mathcal{L}_{\mathrm{ssep}}(\ket{\!\bullet\!\bullet}\bra{\!\bullet\!\bullet\!}) & =0 ~.
\end{align*}
This coincides with the Markov matrix of the classical SSEP. This coincidence also holds for the boundary processes. Thus, the average Q-SSEP dynamics, when reduced to density matrices which are diagonal in the particle number basis, is that of the classical SSEP, as claimed.

As a consequence, the generating function of the steady fluctuations of the classical SSEP occupancies $\tau_j=0,1$ can be expressed as a quantum expectation value w.r.t. the steady averaged Q-SSEP density matrix~:
\beq 
\label{eq:class-SSEP}
\mathbb{E}_\mathrm{ssep}\left[ e^{\sum_j h_j \tau_j} \right] = \mathbb{E}_\infty\Big[ \mathrm{Tr}\big( \rho\, e^{\sum_i h_i \hat n_i}\big)\Big]= \mathrm{Tr}\left( \bar \rho_\infty\, e^{\sum_i h_i \hat n_i}\right) ~,
\eeq
with $\hat n_i :=c^\dag_ic_i$ the quantum number operators, $\bar \rho_\infty:=\mathbb{E}_\infty[\rho]$ the mean Q-SSEP state, averaged w.r.t. the Q-SSEP steady measure denoted $\mathbb{E}_\infty$. In particular, the multi-point correlation functions of the occupation numbers in the classical SSEP coincide with the quantum expectation values of the number operators w.r.t. to steady averaged Q-SSEP density matrix.

To complete this correspondence we need to express the quantum expectation values of the number operators in terms of known data relative to the Q-SSEP steady measure. The latter is constructed by looking at the expectation values of the fermion two-point functions $G_{ij}:=\Tr(\rho\, c^\dag_j c_i)$. It has been shown \cite{BernardJin19} that the leading cumulants of this matrix are those for which the matrix indices are organised along a loop, namely of the form $\mathbb{E}_\infty[ G_{i_1i_n}\cdots G_{i_3i_2}G_{i_2i_1}]^c$. These cumulants scale as $N^{1-n}$ in the scaling limit. We let $\psi^{\#}_n(x_{1},\cdots,x_{n})$ be the scaled cumulants at non-coincident points:
\beq \label{eq:def-CQ}
\psi^{\#}_n(x_{1},\cdots,x_{n}):=\lim_{N\to\infty} N^{n-1}\, \mathbb{E}_\infty\left[ G_{[x_1N][x_nN]}\cdots G_{[x_3N][x_2N]}G_{[x_2N][x_1N]} \right]^c,
\eeq
for $x_k\in[0,1]$ all distincts. The limit is known to exist and to be smooth at non-coincident points. Equations characterising the $\psi^{\#}_n$'s have been written and analysed in \cite{BernardJin20}. They were later shown \cite{Biane22} to be related to free cumulants, as we shall recall below.

Having introduced the main players, we can now state the relation between the non-coincident cumulants in the classical and quantum SSEP, see ref.\cite{BernardJin19}.

\begin{prop} 
For $0<x_k<1$ all different, we have
\beq 
 \label{eq:n-points-class}
\psi_n^\mathrm{ssep}(x_1,\cdots,x_n) = (-1)^{n-1}\sum_{\sigma\in S_n/\mathbb{Z}_n} \psi^{\#}_n(x_{\sigma_1},\cdots,x_{\sigma_n}) .
\eeq
The sum is over all permutations  $\sigma$ modulo cyclic permutations. There are $(n-1)!$ terms in the sum.
\end{prop}

The relation \eqref{eq:n-points-class} essentially follows from Wick's theorem, see ref.\cite{BernardJin19}.
Together with the link between the Q-SSEP cumulants and free probability that we shall describe below, eq.\eqref{eq:n-points-class} is the starting point of the following new construction of the classical SSEP large deviation function.

\subsection{Non-coincident SSEP cumulants from free cumulants}

The aim of this section is, on the one hand, to relate the generating function of non-coincident SSEP cumulants to free probability and, on the other hand, to use this relation to derive a simple integral representation of this generating function.

In order to avoid confusion between the large deviation generating and rate functions as given in the previous SSEP literature \cite{DerridaLarge,Derrida_Review,Mallick_Review} -- namely $F_\mathrm{ssep}[h]$ and $I_\mathrm{ssep}[n]$ defined above in eqs.(\ref{eq:Wh-dev},\ref{eq:Issep-def}) -- and the ones that we shall determine using free cumulant techniques, we shall denote the latter with script letters -- namely $\mathfrak{F}_\mathrm{ssep}[h]$ and $\mathfrak{I}_\mathrm{ssep}[n]$. We shall prove in Section \ref{sec:equivalence} that they (of course) coincide.

The Q-SSEP steady measure, and hence the functions $\psi^{\#}_n$, have been shown to be related to free cumulants~\cite{Biane22}. 

\begin{prop}
Let the interval $[0,1]$ equipped  with the Lebesgue measure, denoted $\mu_L$, be viewed as a probability space.
Let $\mathbb{I}_x=1_{[0,x]}$ be the indicator function of the interval $[0,x]$ with $0<x<1$. 
We have $\mu_L(\mathbb{I}_{x_1}\cdots\mathbb{I}_{x_n})=\min(x_1,\cdots,x_n) =: x_1\wedge \cdots \wedge x_n.$

The loop-expectation values $\psi^\#_n$ are identified as the free cumulants $R_n$ of those random variables with respect to the measure $\mu_L$. Namely~, 
\beq\label{eq:main-free}
\psi^\#_n(x_1,\cdots,x_n) = R_n\big(\mathbb{I}_{x_1},\cdots,\mathbb{I}_{x_n}\big) .
\eeq
\end{prop}

The proof of this fact relied on a combinatorial analysis, see ref.\cite{Biane22} for a set of characterizing equations for the steady cumulants $\psi^\#_n$~\cite{BernardJin20}. See ref.\cite{HruzaBernard22} for an alternative analytic proof based on analysing the time evolution equations of the Q-SSEP cumulants.

As an illustration, we write the first few terms for low values of $n$, using the defining relation between moments and free cumulants \cite{Voiculescu1997,Mingo2017,Speicher2019,Biane2003}. For $n=2,\, 3$, we have~:
\beqs
\psi^\#_2(x,y)&=& x\wedge y - xy ~,\\
\psi^\#_3(x,y,z) &=& (x\wedge y \wedge z) - (x\wedge y)z -  (x\wedge z)y -  (y\wedge z)x + 2 xyz ~,
\eeqs
 If $0<x<y<z<1$, we get
\beqs
\psi^\#_2(x,y)&=& x(1-y) ~,\\
\psi^\#_3(x,y) &=& x(1-2y)(1-z) ~.
\eeqs
For $n\leq 3$ there is no difference between free and standard cumulants. The difference starts at $n\geq 4$. 
For $n=4$, we have, for any order between the points $x_j$:
\beqs
\psi^\#_4(x_1,x_2,x_3,x_4) &=& (x_1\wedge x_2\wedge x_3\wedge x_4) 
- \big((x_1\wedge x_2\wedge x_3)x_4 + \circlearrowleft_{[\mathrm{4\ terms\ in\ total}]}\big) \\
&& +\big((x_1\wedge x_2)x_3x_4 + \circlearrowleft_{[\mathrm{6\ terms\ in\ total}]}\big) - 3x_1x_2x_3x_4\\
&& - \big((x_1\wedge x_2)- x_1x_2)\big)\big((x_3\wedge x_4)- x_3x_4)\big) \\
&& - \big((x_1\wedge x_4)- x_1x_4)\big)\big((x_2\wedge x_3)- x_2x_3)\big)  ~. 
\eeqs
If we choose to order them on the segment $[0,1]$, i.e. $0<x_1<x_2<x_3<x_4<1$, we get:
\beqs
 \psi^\#_4(x_1,x_2,x_3,x_4) &=& x_1(1-3x_2-2x_3+5x_2x_3)(1-x_4) ~,\\
  \psi^\#_4(x_1,x_3,x_4,x_2) &=& x_1(1-3x_2-2x_3+5x_2x_3)(1-x_4) ~,\\
 \psi^\#_4(x_1,x_3,x_2,x_4) &=& x_1(1-4x_2-x_3+5x_2x_3)(1-x_4) ~.
\eeqs
We observe that they depend on the ordering of the points on the line (for $n\geq 4$).
Higher order  cumulants can similarly be computed recursively. Of course, the computation becomes more and more involved and we need to package it.

Let us introduce the generating function of the classical SSEP non-coincident cumulants -- recall the latter are linked to the Q-SSEP expectation values via eq.\eqref{eq:n-points-class}. It is defined by
\beq \label{eq:def-F0}
\mathfrak{F}_0[a;v] := \sum_{n\geq 1} \frac{v^n}{n!}\,\psi^\mathrm{ssep}_n[a],\quad \psi_n[a]:= \int_0^1{\psi}^\mathrm{ssep}_n(x_1,\cdots,x_n) \prod_{k=1}^n a(x_k)dx_k\,,
\eeq
with ${\psi}^\mathrm{ssep}_n$ the scaled non-coincident cumulants. Here, $v$ is simply a counting parameter that we introduced for later convenience. To avoid confusion and to allow for futur comparison with the formula obtained in the SSEP literature, we have used a specific notation for the SSEP large deviation function computed using free cumulant technique. We set $\mathfrak{F}_0[a]:=\mathfrak{F}_0[a;v=1]$. Of course we have $\mathfrak{F}_0[a;v]=\mathfrak{F}_0[av]$.

\begin{lem} The generating function $\mathfrak{F}_0$ of the classical SSEP non-coincident cumulants can be expressed in terms of free cumulants $R_n$ w.r.t. the Lebesgue measure as
\beq \label{eq:F-kappa}
\mathfrak{F}_0[a]= - \sum_{k\geq 1} \frac{(-1)^{n}}{n} R_n(\mathbb{I}_{[a]}) ,
\eeq
with $\mathbb{I}_{[a]}:=\int \!dx\, a(x)\mathbb{I}_x$. Equivalently, $\psi^\mathrm{ssep}_n[a] = (-1)^{n-1}(n-1)! \, R_n(\mathbb{I}_{[a]})$.
\end{lem}

\begin{proof}
This is a direct consequence of the classical-to-quantum SSEP correspondence \eqref{eq:n-points-class} and the multi-linearity of the free cumulants which imply
\beq \label{eq:Cn-Kappa}
\psi_n^\mathrm{ssep}[a] = (-1)^{n-1}(n-1)! \, R_n(\mathbb{I}_{[a]}).
\eeq
\end{proof}

The function $\mathbb{I}_{[a]}$ is a classical variable on $[0,1]$, equipped with the Lebesgue measure, and $R_n(\mathbb{I}_{[a]})$ its $n$-th free cumulants. For a single, hence commuting, variable, there is a simple relation between the generating function of the free cumulants and that of the moments~\cite{Voiculescu1997,Mingo2017,Speicher2019,Biane2003}. This relation goes through the resolvent. We shall now explain how this yields to an efficient way to compute the classical SSEP non-coincident cumulants and how it can be used it to derive a simple integral representation of $\mathfrak{F}_0[a]$.

Let $R_{[a]}(z)$ be the generating function of the free cumulants $R_n(-\mathbb{I}_{[a]})$~:
\beqs 
R_{[a]}(z):=\frac{1}{z}+\sum_{p\geq 1} R_p(-\mathbb{I}_{[a]})z^{p-1} .
\eeqs 
Let $G_{[a]}(z):=\mu_L( \frac{1}{ z+\mathbb{I}_{[a]} })$ be the generating function of the moments $\mu_L( -\mathbb{I}_{[a]}^p )$~:
\beqs
 G_{[a]}(z)=\sum_{p\geq 0} z^{-p-1}\,\mu_L( -\mathbb{I}_{[a]}^p ) .
 \eeqs
 From ref.\cite{Voiculescu1997,Mingo2017,Speicher2019,Biane2003}, these two generating functions are inverse functions, i.e. $R_{[a]}(G_{[a]}(z))=z$, which reads
\beq \label{eq:G-K}
1 + \sum_{p\geq 1} R_p(-\mathbb{I}_{[a]})G_{[a]}(z)^{p} = zG_{[a]}(z).
\eeq
Let $b(x):=-\int_x^1\!dy\, a(y)$, so that $b'(x)=a(x)$ with $b(1)=0$. As a function on $[0,1]$, we have $\mathbb{I}_{[a]}=-b$, since $\mathbb{I}_{[a]}(x)=\int_0^1\!dy\, a(y)\mathbb{I}_{x<y}=\int_x^1\!dy\, a(y)$. Thus, $\varphi( -\mathbb{I}_{[a]}^p) = \int_0^1\!dx\, b(x)^p=:\overline{b^p}$, and
\beq \label{eq:Gb-explicit}
G_{[a]}(z) = \int_0^1 \frac{dx}{z+\mathbb{I}_{[a]}(x)}=  \int_0^1 \frac{dx}{z-b(x)}.
\eeq
We can turn the logic around and view $x$ as a function of $b$ : $x(b)$ is then interpreted as the cumulative probability for $b$, i.e. $dx(b)$ is the probability density for the variable $b$. The fact that $x\in[0,1]$ is then natural.

Formulas (\ref{eq:G-K},\ref{eq:Gb-explicit}) yield a simple recursive way to compute the free cumulants $R_p$ in terms of the moments $\overline{b^p}$ and hence the generating function of the classical SSEP cumulants. We have~:
\beqa \label{eq:free-Ib}
R_1\big(-\mathbb{I}_{[a]}) &=& \overline{b}  ~,\nonumber\\
R_2\big(-\mathbb{I}_{[a]}) &=& \overline{b^2} -\overline{b}^2 ~,\\
R_3\big(-\mathbb{I}_{[a]}) &=& \overline{b^3} - 3\,\overline{b^2}\,\overline{b} + 2\,\overline{b}^3  ~,\nonumber\\
R_4\big(-\mathbb{I}_{[a]}) &=& \overline{b^4} - 4\,\overline{b^3}\, \overline{b} + 10\, \overline{b^2}\, \overline{b}^2 - 2\, \overline{b^2}^2 + 5\,\overline{b}^4 ~,\ \mathrm{etc}.\nonumber
\eeqa
Higher free cumulants can be computed recursively. Using eq.\eqref{eq:Cn-Kappa} this yields the classical SSEP non-coincident cumulants. As a consequence~:

\begin{lem}
Let $b(x):=-\int_x^1\!dy\, a(y)$. We have the following  integral representation of the generating function of non-coincident SSEP cumulants~:
\beq \label{eq:F0-free-ssep}
\mathfrak{F}_0[a] = \int_0^1 \!dx\, \log(z-b(x)) - z + 1,\quad \mathrm{with}\ \int_0^1 \frac{dy}{z-b(y)} =1.
\eeq 
\end{lem}

\begin{proof}
Recall that $\mathfrak{F}_0[a;v]=\mathfrak{F}_0[av]$. By definition, eq.\eqref{eq:def-F0}, we have
\[ 
v\partial_v \mathfrak{F}_0[a;v] =  - \sum_{k\geq 1} v^{n} R_n(-\mathbb{I}_{[a]}) = 1- v R_{[a]}(v) ,
\]
with $R_{[a]}(z):=\frac{1}{z}+\sum_{p\geq 1} \kappa_p(-\mathbb{I}_{[a]})z^{p-1}$ as above. Thus, the relation $R_{[a]}(G_{[a]}(z))=z$ becomes
\beq \label{eq:for-F}
1 - v\partial_v \mathfrak{F}_0[a;v] = vz, \quad \mathrm{with}\ v=\int_0^1 \frac{dy}{z-b(y)}.
\eeq
We simply have to check that the function \eqref{eq:F0-free-ssep} is indeed solution of the above equation, with the appropriate boundary condition. Assuming $\mathfrak{F}_0[a]$ given by the r.h.s. of \eqref{eq:F0-free-ssep}, we have 
\[
\mathfrak{F}_0[a;v] = \int_0^1 \!dx\, \log[v(z-b(x))] - vz + 1,
\]
with $\int_0^1 \frac{dy}{z-b(y)} =v$. Computing its derivative with respect to $v$, we get (using that $z=z[b,v]$ is actually a function of $v$ and $b$)
\[
v\partial_v \mathfrak{F}_0[a;v] = 1-vz + v(\frac{\partial z}{\partial v})\big(\int_0^1\frac{dx}{z-b(x)} -v\big) .
\]
The last term vanish by definition of $z=z[b,v]$ and we get $v\partial_v \mathfrak{F}_0[a;v] = 1-vz$, as required. It is easy to check that the function \eqref{eq:F0-free-ssep} has the appropriate behavior at small $v$.
\end{proof}

This representation can also be used to recursively compute $\mathfrak{F}_0[a;v]$. Eq.\eqref{eq:for-F} can alternatively be written as
\beq \label{eq:W-bis}
v\partial_v\mathfrak{F}_0[a;v] = 1 - vz,\quad \mathrm{with} \quad v= G_{[a]}(z) ~.
\eeq
The last relation, $v=G_{[a]}(z)$, determines $z$ as a function of $v$, recursively~: 
\beqs
vz(v) &=& 1 + \overline{b}\,v +(\overline{b^2}-\overline{b}^2)\,v^2 + (\overline{b^3} - 3\, \overline{b^2}\, \overline{b} + 2\,\overline{b}^3)\,v^3\\
&& + (\overline{b^4} - 4\,\overline{b^3}\, \overline{b} + 10\, \overline{b^2}\, \overline{b}^2 - 2\, \overline{b^2}^2 + 5\,\overline{b}^4 )\,v^4 
+\cdots ~,
\eeqs
This is of course the generating function of the free cumulants of $b$, see eq.\eqref{eq:free-Ib}. 
This is a very efficient way to compute the multi-point SSEP cumulants at non-coincident points.

\subsection{The classical SSEP large deviation function from free probability}

Once the generating function $\mathfrak{F}_0[a]$ of non-coincident SSEP cumulants has been identified, the SSEP large deviation generating function $\mathfrak{F}_\mathrm{ssep}[h]$ can be computed using the variational principle established in the previous Sections \ref{sec:Combinatoire} \& \ref{sec:Feynman}~:
\beq \label{eq:F-ssep}
\mathfrak{F}_\mathrm{ssep}[h]= \max_{g(\cdot);q(\cdot)} \left[\int_0^1 \!\!\! dx\big[ \log\big(1+g(x)(e^{h(x)}-1)\big) - q(x)g(x)\big] + \mathfrak{F}_0[q]\right] ,
\eeq
with $\mathfrak{F}_0[q]$ defined in eq.\eqref{eq:F0-free-ssep}.
Since the rate function $\mathfrak{I}_\mathrm{ssep}$ is the Legendre transform of the large deviation generating function, we have

\begin{coro}
\begin{equation} \label{eq:rate-ssep}
\mathfrak{I}_\mathrm{ssep}[n]=\max_{g(\cdot), q(\cdot)} \left( \int_0^1\!\! dx \, \left[ n(x)\log\big(\frac{n(x)}{g(x)}\big)+(1-n(x))\log\big(\frac{1-n(x)}{1-g(x)}\big)+q(x)g(x) \right]-\mathfrak{F}_0[q]\right).
\end{equation}
\end{coro}

\begin{proof}
Using eq.\eqref{eq:F-ssep} and $\mathfrak{I}_\mathrm{ssep}[n]=\max_{h(\cdot)} \left(\int_0^1 \!\! dx\, h(x)n(x) - \mathfrak{F}_\mathrm{ssep}[h]\right)$,  we write
\begin{equation} \label{eq:rate-bis}
\mathfrak{I}_\mathrm{ssep}[n]=\max_{h(\cdot)} \left(\int_0^1  \!\! dx\, \left[ h(x) n(x) - \log\big(1+g(x)e(x)\big) + q(x)g(x)\right] - \mathfrak{F}_0[q] \right),
\end{equation}
with $e(x)=e^{h(x)}-1$.
Here $g$ and $q$ are determined as the functions for which $\mathfrak{F}_\mathrm{ssep}[h]$ takes its maximal value,
\begin{align}\label{eq:condition}
q(x)=\frac{e(x)}{1+g(x)e(x)} \quad,\quad 
g(x)=\frac{\delta \mathfrak{F}_0[q]}{\delta q(x)}.
\end{align}
The maximum of (\ref{eq:rate-bis}) is attained for
\begin{equation*} 
h(x)=\log\left(\frac{n(x)(1-g(x))}{g(x)(1-n(x))}\right),
\end{equation*}
from which we get 
\beq \label{eq:ngq}
1+g(x)e(x)=\frac{1-g(x)}{1-n(x)},\quad q(x)=\frac{n(x)}{g(x)}-\frac{1-n(x)}{1-g(x)} .
\eeq
 Inserting this back into the expression the rate function yields
\begin{equation*}
\mathfrak{I}_\mathrm{ssep}[n]=\int_0^1 dx \, \left[n \log\big(\frac{n}{g}\big)+(n-1)\log\big(\frac{1-g}{1-n}\big)+qg \right]-\mathfrak{F}_0[q].
\end{equation*}
The two conditions \eqref{eq:condition} or \eqref{eq:ngq} can be relaxed in writing this expression as a maximization problem (and simplifying the expression) as in eq.\eqref{eq:rate-ssep}.
Indeed, one can check that the extremization condition \eqref{eq:rate-ssep} yields the same conditions for $g$ and $q$ as in \eqref{eq:condition} or \eqref{eq:ngq}. 
\end{proof}

\subsection{Equivalence with the previously known formulation}
\label{sec:equivalence}

Finally, we present an explicit check that our new formula \eqref{eq:F-ssep} for the SSEP large deviation function is identical to the previously known formula \eqref{eq:Wh-int}. 

To prove this equivalence, we first formulate differently, but equivalently, the variational problem \eqref{eq:Wh-int}, as follows~:
\beqa \label{eq:Wh-new}
F_\mathrm{ssep}[h] &=& \max_{g(\cdot);f(\cdot)} \widehat F[h;f,g],
\quad \widehat F[h;f,g]:= \int_0^1 \!\!\! dx\big[ \log\big(1+g(x)\, e(x)\big) - f(x)g(x)\big] + V[f] ,
\nonumber
\eeqa
where the functional $V[f]$ be defined by
\beq \label{eq:def-Vh}
V[f] := \int_0^1 dx\, \log\big(w-\ell(x)\big) - w + 1,\quad \mathrm{with}\ \int_0^1 \frac{dy}{w-\ell(y)} =1,
\eeq
with $\ell(x):=-\int_x^1 dy f(y)$, so that $\ell'(x)=f(x)$ with $\ell(1)=0$. We view $w$ as a function of $\ell$ through the constraint $\int_0^1 \frac{dy}{w-\ell(y)} =1$. 

To prove the equivalence between the two variational problems, we first have to compute the functional derivative of $V[f]$. Chain rule implies
\[ 
\frac{\delta V[f]}{\delta f(x)} = -\int_0^1 (\frac{\delta \ell(y)}{\delta f(x)})( \frac{dy}{w-\ell(y)}) + \big(\frac{\delta w}{\delta f(x)}\big)\,\big(\int_0^1 \frac{dy}{w-\ell(y)} -1\big).
\]
The second term vanish due to the relation $\int_0^1 \frac{dy}{w-\ell(y)} =1$. The definition of $\ell$ as $\ell(x):=-\int_x^1 dy f(y)$ implies $\frac{\delta \ell(y)}{\delta f(x)}=-\mathbb{I}_{\{x>y\}}$. Thus
\beq \label{eq:Vh-derive}
\frac{\delta V[f]}{\delta f(x)} = \int_0^x \frac{dy}{w-\ell(y)}.
\eeq
The extremization conditions \eqref{eq:Wh-int} read
\beqs
f(x) =  \frac{e(x)}{1+e(x)g(x)} ,\quad
g(x) = \frac{\delta V[f]}{\delta f(x)} = \int_0^x \frac{dy}{w-\ell(y)}.
\eeqs
The relation $\int_0^1 \frac{dy}{w-\ell(y)} =1$ implies the boundary conditions $g(1)=1$.
The last condition is equivalent to $1/g'(x)= w-\ell(x)$ with $g(0)=0$, and hence to $(1/g'(x))'= -\ell'(x)=-f(x)$, with $f(x)=\frac{e(x)}{1+e(x)g(x)}$, which is then equivalent to \eqref{eq:Wh-extrem}.

The last step consists in verifying that the extremum value coincide. Thanks to the extremum conditions, written as $g'(x)(w-\ell(x))=1$, and the boundary conditions on $g$ and $\ell$, we have
\[
\int_0^1 \!dx\, f(x)g(x)= \int_0^1 \!dx\, \ell'(x)g(x)= -\int_0^1 \!dx\, \ell(x)g'(x) = 1-w,
\]
so that 
\[
\widehat F[h;f,g]\vert_\mathrm{ext} - F[h;g]\vert_\mathrm{ext} = -\int_0^1 \!dx\, f(x)g(x) - w +1 =0
\]

It is now clear that the two extremization problems \eqref{eq:Wh-int} and \eqref{eq:F-ssep} are equivalent, with the correspondence is $w\leadsto z$, $\ell\leadsto -\mathbb{I}_{[a]}$, so that $\mathfrak{F}_\mathrm{ssep}[h] = F_\mathrm{ssep}[h]$.

\vspace{1cm}

\appendix

\centerline{\textbf{\LARGE{Appendices}}}

\section{Formalities}\label{appsec:F}

\subsection{Formal Power Series}\label{appssec:FPS}

If $R$ is any commutative ring with unit, $S$ is an arbitrary index set and $\lambda_\mybullet:=(\lambda_s)_{s\in S}$ a collection of variables, we denote by $R[\lambda_\mybullet]$, the ring of polynomials and by $R(\lambda_\mybullet)$ the ring of formal power series in the variables $\lambda_\mybullet$ with coefficients in $R$. We let $R(\lambda_\mybullet)_{\geq 1}$ denote the ideal in $R(\lambda_\mybullet)$ of formal powers series with vanishing constant coefficient. The rings $R[\lambda_\mybullet]$ and $R(\lambda_\mybullet)$ are again commutative rings with unit, so if $J$ is a new arbitrary index set and $z_\mybullet:=(z_i)_{i\in J}$, ${\overline z}_\mybullet:=({\overline z}_i)_{i\in J}$ are new variables we may consider for instance $R[z_\mybullet,{\overline z}_\mybullet](\lambda_\mybullet)$, the ring of formal power series in the variables $\lambda_\mybullet$ with coefficients in the ring $R[z_\mybullet,{\overline z}_\mybullet]$ of polynomials in $z_\mybullet,{\overline z}_\mybullet$. We note that $R[z_\mybullet,{\overline z}_\mybullet](\lambda_\mybullet)\subset R(z_\mybullet,{\overline z}_\mybullet,\lambda_\mybullet)$. Notice that in $R[z_\mybullet,{\overline z}_\mybullet](\lambda_\mybullet)$ the ``constant'' coefficient is now a polynomial in $R[z_\mybullet,{\overline z}_\mybullet]$ and this polynomial vanishes for members of $R[z_\mybullet,{\overline z}_\mybullet](\lambda_\mybullet)_{\geq 1}$. If $F$ is any formal power series in one variable and $L_{\lambda_\mybullet}(z_\mybullet,{\overline z}_\mybullet)\in R[z_\mybullet,{\overline z}_\mybullet](\lambda_\mybullet)_{\geq 1}$ then the composition $F\circ L_{\lambda_\mybullet}(z_\mybullet,{\overline z}_\mybullet)$ is well-defined as an element of $R[z_\mybullet,{\overline z}_\mybullet](\lambda_\mybullet)$. For instance if $A_{\lambda_\mybullet}(z_\mybullet,{\overline z}_\mybullet)\in R[z_\mybullet,{\overline z}_\mybullet](\lambda_\mybullet)$ then $A_{\lambda_\mybullet}(z_\mybullet,{\overline z}_\mybullet) e^{L_{\lambda_\mybullet}(z_\mybullet,{\overline z}_\mybullet)} \in R[z_\mybullet,{\overline z}_\mybullet](\lambda_\mybullet)$.

\subsection{Formal Gaussian Integrals}\label{appssec:FGI}

Form now on, the ground ring $R$ is a field of characteristic $0$, say $\mathbb{R}$. The restriction to characteristic $0$ is mostly for convenience, it avoids for instance to deal explicitly with divided powers.   

Our first aim is to make sense of 
\[ \int \left(\prod_{i\in J} \frac{d{\overline z}_i\wedge dz_i}{2i\pi\hbar} \exp (- z_i {\overline z}_i/\hbar) \right) A_{\lambda_\mybullet}(z_\mybullet,{\overline z}_\mybullet) \]
where $\hbar$ is yet another formal variable and $A_{\lambda_\mybullet}(z_\mybullet,{\overline z}_\mybullet)\in R[z_\mybullet,{\overline z}_\mybullet](\lambda_\mybullet)$. The result of integration will be a element of $R(\hbar,\hbar^{-1},\lambda_\mybullet)$ and in fact $A_{\lambda_\mybullet}(z_\mybullet,{\overline z}_\mybullet)$ may contain $\hbar$ explicitly (the notation is already heavy enough). It is useful to notice that contrary to the other formal variables involved, $\hbar$ can be specialized to a numerical value without impact on most of the discussions that follow (the semi-classical expansion below being an important exception) and then the result of integration is in $R(\lambda_\mybullet)$. 
We define the integral by term by term integration of the $\lambda_\mybullet$-formal power series expansion of $A_{\lambda_\mybullet}(z_\mybullet,{\overline z}_\mybullet)$ so we are left with the task of defining
\[ \int \left(\prod_{i\in J} \frac{d{\overline z}_i\wedge dz_i}{2i\pi\hbar} \exp (- z_i {\overline z}_i/\hbar) \right) P(z_\mybullet,{\overline z}_\mybullet) \] for $P(z_\mybullet,{\overline z}_\mybullet) \in R[z_\mybullet,{\overline z}_\mybullet]$. Imposing linearity, it is enough to deal with monomials in $R[z_\mybullet,{\overline z}_\mybullet]$ i.e. expressions of the form $\prod_{i\in J} z_i^{m_i} {\overline z}_i^{{\overline m}_i}=:z_\mybullet^{m_\mybullet} {\overline z}_\mybullet^{{\overline m}_\mybullet}$ where $m_i,{\overline m}_i\in \mathbb{N}$ and $\sum_{i\in J} m_i+{\overline m}_i$, called the degree of the monomial, is finite. For such a monomial we set 
\[ \int \left(\prod_{i\in J} \frac{d{\overline z}_i\wedge dz_i}{2i\pi\hbar} \exp (- z_i {\overline z}_i/\hbar) \right) \prod_{i\in J} z_i^{m_i} {\overline z}_i^{{\overline m}_i}:=  \prod_{i\in J} \delta_{m_i,{\overline m}_i}\hbar^{m_i} m_i! ,\]
a formula copied from the honest integral over the complex plane
\[ \int_{\mathbb{C}} \frac{d{\overline z}\wedge dz}{2i\pi\hbar} \exp (- z {\overline z}/\hbar)  z^m {\overline z}^{\overline m}=\delta_{m,{\overline m}}\hbar^{m} m! ,\]
which holds for $m,{\overline m}\in \mathbb{N}$ and $\hbar$ a complex number with strictly positive real part.

Our main interest lies in the computation of
\[ \int \left(\prod_{i\in J} \frac{d{\overline z}_i\wedge dz_i}{2i\pi\hbar} \exp (- z_i {\overline z}_i/\hbar) \right)  \exp L_{\lambda_\mybullet}(z_\mybullet,{\overline z}_\mybullet)/\hbar \]
where $L_{\lambda_\mybullet}(z_\mybullet,{\overline z}_\mybullet)\in  R[z_\mybullet,{\overline z}_\mybullet](\lambda_\mybullet)_{\geq 1}$ and $\hbar$ is yet another formal variable. As $\exp L_{\lambda_\mybullet}(z_\mybullet,{\overline z}_\mybullet)/\hbar$ belongs to $R[z_\mybullet,{\overline z}_\mybullet](\lambda_\mybullet)$ this is really a special case of the previous discussion.

As a warming exercise, the reader is invited to check that, as a formal integral where $z,{\overline z},u$ and ${\overline v}$ are formal variables, 
\[ \int \frac{d{\overline z}\wedge dz}{2i\pi\hbar} \exp (- (z+u)({\overline z}+{\overline v})/\hbar)=1.\]
This simple identity, which can be read as translation invariance (independently over $z, {\overline z}$ so its validity for its honest integral avatar does not reduce to translation invariance of the Lebesgue measure), plays an important role in the manipulation of (formal) Gaussian integrals. The reader should also check the corollaries
\[ \int \frac{d{\overline z}\wedge dz}{2i\pi\hbar} \exp (- z{\overline z}/\hbar)\exp (- ({\overline v}z/ +u{\overline z})/\hbar)=e^{u {\overline v}/\hbar},\]
\[ \int \frac{d{\overline z}\wedge dz}{2i\pi\hbar} \exp (- z{\overline z}/\hbar)\exp (((\overline{\nu}-{\overline v})z +(\mu-u){\overline z}+{\overline \nu}u+\mu {\overline v}-u{\overline v})/\hbar)=e^{\mu {\overline \nu}/\hbar},\]
whose expansion in the new formal variables $\mu,{\overline \nu}$ yields a more general version of translation invariance
\[  \int \frac{d{\overline z}\wedge dz}{2i\pi\hbar} \exp (- (z+u)({\overline z}+{\overline v})/\hbar) (z+u)^m ({\overline z}+{\overline v})^{{\overline m}}= \delta_{m,{\overline m}}\hbar^{m} m!,\]
which also plays an important role in what follows.

\subsection{Semi-classical expansion}\label{appssec:SCE}

We introduce a collection of  $u_\mybullet:=(u_i)_{i\in J}$, ${\overline v}_\mybullet:=({\overline v}_i)_{i\in J}$ of formal variables and use translation invariance:
\begin{eqnarray*} \int \left(\prod_{i\in J} \frac{d{\overline z}_i\wedge dz_i}{2i\pi\hbar} \exp (- z_i {\overline z}_i/\hbar) \right) \exp L_{\lambda_\mybullet}(z_\mybullet,{\overline z}_\mybullet)/\hbar & = & \\
& & \hspace{-7cm} \int \left(\prod_{i\in J} \frac{d{\overline z}_i\wedge dz_i}{2i\pi\hbar} \exp (- (z_i+u_i)({\overline z}_i+{\overline v}_i)/\hbar) \right)  \exp L_{\lambda_\mybullet}(z_\mybullet+u_\mybullet,{\overline z}_\mybullet+{\overline v}_\mybullet)/\hbar.\end{eqnarray*}
We infer that if $U_\mybullet:=(U_i)_{i\in J}$, ${\overline V}_\mybullet:=(V_i)_{i\in J}$ are arbitrary elements of $R(\lambda_\mybullet)_{\geq 1}$ then                 
\begin{eqnarray*} \int \left(\prod_{i\in J} \frac{d{\overline z}_i\wedge dz_i}{2i\pi\hbar} \exp (- z_i {\overline z}_i/\hbar) \right) \exp L_{\lambda_\mybullet}(z_\mybullet,{\overline z}_\mybullet)/\hbar & = & \\
& & \hspace{-7cm} \int \left(\prod_{i\in J} \frac{d{\overline z}_i\wedge dz_i}{2i\pi\hbar} \exp (- (z_i+U_i)({\overline z}_i+{\overline V}_i)/\hbar) \right)  \exp L_{\lambda_\mybullet}(z_\mybullet+U_\mybullet,{\overline z}_\mybullet+{\overline V}_\mybullet)/\hbar.\end{eqnarray*}
We claim that for any given $B$ there are unique members $U_\mybullet^*$, ${\overline V}_\mybullet^*$ in $R(\lambda_\mybullet)_{\geq 1}$ such that
\[ \sum_{i\in J} (z_i+U_i^*) ({\overline z}_i+{\overline V}_i^*)-L_{\lambda_\mybullet}(z_\mybullet+U_\mybullet^*,{\overline z}_\mybullet+{\overline V}_\mybullet^*)\]
has an extremum in $z_\mybullet$ and  ${\overline z}_\mybullet$ at $0$. Indeed the extremum equations are
\[ {\overline V}_i=\frac{\partial L_{\lambda_\mybullet}}{\partial z_i} (U_\mybullet,{\overline V}_\mybullet) \qquad U_i=\frac{\partial L_{\lambda_\mybullet}}{\partial {\overline z}_i} (U_\mybullet,{\overline V}_\mybullet)\]
for $i\in J$, and existence/uniqueness of  $(U_\mybullet^*,{\overline V}_\mybullet^*)$ follow from a tedious but straightforward recursive argument on the degrees in the $(z_\mybullet^*,{\overline z}_\mybullet^*)$ of $L_{\lambda_\mybullet}$.

Set $F^*:= -\sum_{i\in J} U_i^*{\overline V}_i^*+L_{\lambda_\mybullet}(U_\mybullet^*,{\overline V}_\mybullet^*)$, a member of $R(\lambda_\mybullet)_{\geq 1}$. Then
\[ -\sum_{i\in J} (z_i+U_i^*) ({\overline z}_i+{\overline V}_i^*)+L_{\lambda_\mybullet}(z_\mybullet+U_\mybullet^*,{\overline z}_\mybullet+{\overline V}_\mybullet^*)-F_{\lambda_\mybullet}^*+\sum_{i\in J} z_i{\overline z}_i=: L_{\lambda_\mybullet}^*(z_\mybullet,{\overline z}_\mybullet)\] belongs to $R[z_\mybullet,{\overline z}_\mybullet](\lambda_\mybullet)_{\geq 1}$ (this fact concerns the $\lambda_\mybullet$-expansion) and involves only terms of degree $\geq 2$ in $(z_\mybullet,{\overline z}_\mybullet)$. 
Consequently, a simple power counting argument shows that
\[ \int \left(\prod_{i\in J} \frac{d{\overline z}_i\wedge dz_i}{2i\pi\hbar} \exp (- z_i {\overline z}_i/\hbar) \right) \exp L_{\lambda_\mybullet}^*(z_\mybullet,{\overline z}_\mybullet)/\hbar \] 
belongs to $R(\hbar,\lambda_\mybullet)$ (no $\hbar^{-1}$ involved) and that the constant term is $1$.
Putting everything together leads to
\[ \int \left(\prod_{i\in J} \frac{d{\overline z}_i\wedge dz_i}{2i\pi\hbar} \exp (- z_i {\overline z}_i/\hbar) \right) \exp L_{\lambda_\mybullet}(z_\mybullet,{\overline z}_\mybullet)/\hbar =\exp F_{\lambda_\mybullet}^*/\hbar +\sum_{n\in \mathbb{N}} \hbar^n F_{\lambda_\mybullet}^{(n)} \]
where each $F_{\lambda_\mybullet}^{(n)}$ belongs to $R(\lambda_\mybullet)_{\geq 1}$.
This $\hbar$-expansion is the formal version of the saddle point expansion. We shall soon see a diagrammatic interpretation of this result.

\section{Feynman Graphs and Rules}\label{appsec:FGR} 

We focus for a while on the case when the index set, $J$ for the variables $(z_\mybullet,{\overline z}_\mybullet)$ is fixed. We denote by $S(J)$ the set indexing monomials in $(z_\mybullet,{\overline z}_\mybullet)$, i.e. 
\[ S(J):=\{(m_i,{\overline m}_i)_{i\in J}, \, m_i,{\overline m}_i\in \mathbb{N} \text{ for } i\in J,\, \sum_{i\in J} m_i+{\overline m}_i <+\infty \}.\]
The variables indexed by $S(J)$ are denoted by $\Lambda_\mybullet:=(\Lambda_{m_\mybullet,{\overline m}_\mybullet})_{(m_\mybullet,{\overline m}_\mybullet)\in S(J)}$ and we concentrate on the computation of
\[ \int \left(\prod_{i\in J} \frac{d{\overline z}_i\wedge dz_i}{2i\pi\hbar} \exp (- z_i {\overline z}_i/\hbar) \right) \exp L(J)_{\Lambda_\mybullet}(z_\mybullet,{\overline z}_\mybullet)/\hbar, \]
where
 \[ L(J)_{\Lambda_\mybullet}(z_\mybullet,{\overline z}_\mybullet):=\sum_{(m_\mybullet,{\overline m}_\mybullet)\in S(J)} \Lambda_{m_\mybullet,{\overline m}_\mybullet} \prod_{i\in J} \frac{z_i^{m_i}}{m_i!} \frac{{\overline z}_i^{{\overline m}_i}}{{\overline m}_i!}\in R[z_\mybullet,{\overline z}_\mybullet](\Lambda_\mybullet)_{\geq 1}.\]
The appearance of factorials in the denominator (leading to so-called divided powers) will simplify the forthcoming formul\ae. 

This is a kind of master integral. Indeed, if $B$ is an arbitrary set indexing variables $\lambda_\mybullet:=(\lambda_s)_{s\in B}$ any $L_{\lambda_\mybullet}(z_\mybullet,{\overline z}_\mybullet)\in R[z_\mybullet,{\overline z}_\mybullet](\lambda_\mybullet)_{\geq 1}$  can by expanded in monomials:
\[ L_{\lambda_\mybullet}(z_\mybullet,{\overline z}_\mybullet)=\sum_{(m_\mybullet,{\overline m}_\mybullet)\in S(J)} \Lambda(B)_{m_\mybullet,{\overline m}_\mybullet} \prod_{i\in J} \frac{z_i^{m_i}}{m_i!} \frac{{\overline z}_i^{{\overline m}_i}}{{\overline m}_i!}\]
where each $\Lambda(B)_{m_\mybullet,{\overline m}_\mybullet}$ belongs to $R(\lambda_\mybullet)_{\geq 1}$. Notice however that the formal powers series $\Lambda(B)_{m_\mybullet,{\overline m}_\mybullet}$ are not arbitrary: a given $\lambda_\mybullet$ appears in only finitely many $\Lambda(B)_{m_\mybullet,{\overline m}_\mybullet}$ so that its coefficient in the $\lambda_\mybullet$ expansion of $L_{\lambda_\mybullet}(z_\mybullet,{\overline z}_\mybullet)$ is indeed a polynomial in $z_\mybullet,{\overline z}_\mybullet$. Anyway, 
\[ \int \left(\prod_{i\in J} \frac{d{\overline z}_i\wedge dz_i}{2i\pi\hbar} \exp (- z_i {\overline z}_i/\hbar) \right) \exp L_{\Lambda_\mybullet}(z_\mybullet,{\overline z}_\mybullet)/\hbar \]
is recovered from 
\[ \int \left(\prod_{i\in J} \frac{d{\overline z}_i\wedge dz_i}{2i\pi\hbar} \exp (- z_i {\overline z}_i/\hbar) \right) \exp L(J)_{\Lambda_\mybullet}(z_\mybullet,{\overline z}_\mybullet)/\hbar\]
by substitution, for each $(m_\mybullet,{\overline m}_\mybullet)\in S(J)$ of the formal power series $\Lambda(B)_{m_\mybullet,{\overline m}_\mybullet}$ for the formal variable $\Lambda_{m_\mybullet,{\overline m}_\mybullet}$.

We turn to graphical rules allowing the computation (in principle) of the master integral. Consider a given $\Lambda_\mybullet$-monomial in the expansion of $\exp L(J)_{\Lambda_\mybullet}(z_\mybullet,{\overline z}_\mybullet)/\hbar$. It comes with a coefficient which is proportional (the coefficient is in  $R$) to a $(z_\mybullet,{\overline z}_\mybullet)$ monomial, and this monomial survives integration (gives a non-zero contribution) if and only if for each $i\in J$ the power of $z_i$ and ${\overline z}_i$ are equal, say $n_i$. Then its integral is $\prod_{i\in J} n_i!$.

For $(m_\mybullet,{\overline m}_\mybullet)\in S(J)$ we represent $\Lambda_{m_\mybullet,{\overline m}_\mybullet} \prod_{i\in J} z_i^{m_i}{\overline z}_i^{{\overline m}_i}$ as a vertex with $m_i$ outgoing edges and ${\overline m}_i$ incoming edges carrying the symbol $i$. Thus a term in the expansion of $\exp L(J)_{\Lambda_\mybullet}(z_\mybullet,{\overline z}_\mybullet)/\hbar$ is represented as a collection of vertices with pending edges. We represent the integration as a black box in which all the edges meet. If the number of incoming edges and outgoing edges carrying the symbol $i\in J$ are not equal, we get $0$, and if both equal $n_i$, we get a factor $n_i!$. This is precisely the number of ways to pair the incoming edges of type $i$ to the outgoing edges of type $i$, that is, the number of ways to organize the inside of the black box so that no pending edge remains, and for each such choice, opening the black box reveals an oriented graph, in which each edges carries a type $i\in J$. Conversely, given such a graph, breaking each edge in two pending edges (keeping track of the type and the orientation) one reconstructs a term in the expansion of $\exp L(J)_{\Lambda_\mybullet}(z_\mybullet,{\overline z}_\mybullet)/\hbar$. The $m_i$ outgoing edges of type $i$ associated to $\Lambda_{m_\mybullet,{\overline m}_\mybullet} \prod_{i\in J} z_i^{m_i}{\overline z}_i^{{\overline m}_i}$ play an equivalent role so permuting them does not change the wiring, so the divided power factor $m_i!$, together with its cousin ${\overline m}_i!$, are absorbed when only the graph is considered. In the same way, permuting the vertices associated to the same $(m_\mybullet,{\overline m}_\mybullet)$ can be compensated by a permutation of their pending edges to preserve the wiring, so that when only the graph is considered the factorials in the expansion of the exponential are reabsorbed. All in all, going from the expansion of $\exp L(J)_{\Lambda_\mybullet}(z_\mybullet,{\overline z}_\mybullet)/\hbar$ to the graphical representation removes all combinatorial factors. Or almost so: it may happen that performing permutations simultaneously for all vertices associated to the same $(m_\mybullet,{\overline m}_\mybullet)$ and all the pending edges of the same type, whether pending at the same vertex or at different vertices, leads to the same graph. The construction of the graph from the expansion suggest the required modification: \\
-- Given the graph $G$, label all the vertices, break each edge in two pending edges (keeping track of the type and the orientation) and label all the pending edges by their type, orientation, and an additional label (so that all pending edges have distinct labels).\\
-- Build two matrices. An incidence matrix whose rows are indexed by the (labelled) vertices and whose columns are indexed by the (labelled) pending edges, with a $1$ at the intersection of a column and a row if the corresponding edge pends to the corresponding vertex in $G$ and $0$ else. An adjacency matrix whose rows are indexed by (labelled) outgoing edges and whose columns are indexed by the (labelled) incoming edges, with a $1$ at the intersection of a column and a row if the corresponding pending edges join to make an edge of $G$ and $0$ else.\\
-- Consider the group which is the direct product of permutations of the vertices, incoming pending edges of a given type and outgoing pending edges of a given type respectively. This groups acts on the incidence and the adjacency matrices, and the automorphism group $\text{Aut}\, G$ of $G$ is the subgroup fixing the two matrices.\\
-- The combinatorial factor that remains when going from the expansion of $\exp L(J)_{\Lambda_\mybullet}(z_\mybullet,{\overline z}_\mybullet)/\hbar$ to the graphical representation is $1/ \# \text{Aut}\, G$, the inverse of the cardinal of $\text{Aut}\, G$.

To summarize, we have obtained the following result:
\[ \int \left(\prod_{i\in J} \frac{d{\overline z}_i\wedge dz_i}{2i\pi\hbar} \exp (- z_i {\overline z}_i/\hbar) \right) \exp L(J)_{\Lambda_\mybullet}(z_\mybullet,{\overline z}_\mybullet)/\hbar = \sum_G w(G) \]
where the sum is over all oriented graphs whose edges carry a type $i\in J$ and the weight $w(G)$ of a graph is computed as follows:\\
-- The edges incident at a vertex define a vertex type $(m_\mybullet,{\overline m}_\mybullet) \in S(J)$ counting how many incoming and outgoing edges of each type are incident, leading to a factor $\Lambda_{m_\mybullet,{\overline m}_\mybullet}\hbar^{-1}$ in $w(G)$.\\
-- Each edge contributes a factor $\hbar$ in $w(G)$.\\
-- There is an overall factor $1/ \# \text{Aut}\, G$ in $w(G)$.

We give a few examples. For concreteness we take $J$ to be the standard alphabet $J:=\{a,b,\cdots,y,z\}$ (of which we use only a subset in the examples!). \\
-- Example: A vertex with $3$ outgoing and $2$ incoming edges of type $f$, and an incoming edge of type $u$. It is convenient to use a compact notation, additive or multiplicative for instance, and denote the corresponding coupling by $\Lambda_{3f,2{\overline f}+{\overline u}}$ (additive, chosen below) or $\Lambda_{f^3{\overline f}^2{\overline u}}$ (multiplicative):
\begin{center}
\includegraphics[width=.15\linewidth]{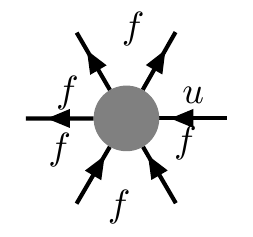}
\end{center}
-- Example: A diagram
\begin{center}
\includegraphics[width=.4\linewidth]{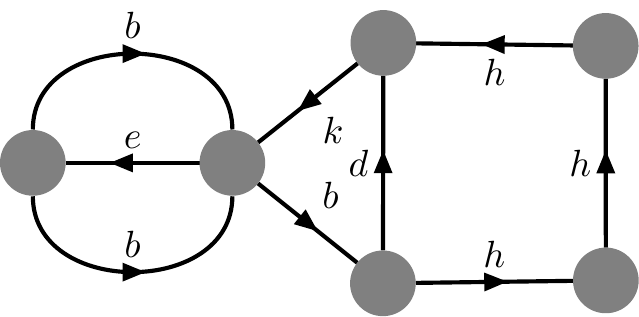}
\end{center}
with weight
\[ \frac{1}{2!}\Lambda_{2b,{\overline e}} \Lambda_{b+e,2{\overline b}+{\overline k}} \Lambda_{k,{\overline d}}\Lambda_{d+h,{\overline b}}\Lambda_{h,{\overline h}}^2.\]
The symmetry number $1/2!$ comes from the two equivalent outgoing type $b$ edges at the vertex on the left.

Letting  $n_v(G)$, $n_e(G)$, $n_l(G)$ and $n_c(G)$ denote respectively the number of vertices, edges, loops and connected components of $G$ we note that the overall power of $\hbar$ in $w(G)$ is $n_e(G)-n_v(G)$, which is also $n_l(G)-n_c(G)$ by a standard topological relation. At this point, we may pause to ask why it is worth to bother going from the plain ``black box expansion'' version of the master integral to the ``graphical'' version: after all, one black box gives rise to a whole family of graphs. One of the advantages of the graphical representation is that it behaves very nicely from the combinatorial viewpoint: the weight $w(G)$ factors nicely over connected components, so that
\[ \int \left(\prod_{i\in J} \frac{d{\overline z}_i\wedge dz_i}{2i\pi\hbar} \exp (- z_i {\overline z}_i/\hbar) \right) \exp L(J)_{\Lambda_\mybullet}(z_\mybullet,{\overline z}_\mybullet)/\hbar = exp \sum_G w(G) \]
where now the sum is over all connected oriented graphs whose edges carry a type $i\in J$, the weight being computed as before. The power of $\hbar$ in $w(G)$ is simply $n_l(G)-1$ if $G$ is connected, and comparing with the semi-classical expansion in \autoref{appssec:SCE} we infer that for the master integral the $(z_\mybullet,{\overline z}_\mybullet)$-extremum $F(J)_{\Lambda_\mybullet}^*$ of $\sum_{i\in J} z_i{\overline z}_i-L(J)_{\Lambda_\mybullet}(z_\mybullet,{\overline z}_\mybullet)$, and the corrections $F(J)_{\Lambda_\mybullet}^{(n)}$, $n=0,1,\cdots$ are given by
\[ F(J)_{\Lambda_\mybullet}^{(n)}=\hbar^{-n}\sum_G w(G) \text{ for } n \geq -1,\] 
where the sum is over connected oriented graphs with $n+1$ loops whose edges carry a type $i\in J$ and $*$ is interpreted as $(-1)$.

\subsection{A special case}\label{appssec:SC}

We turn to a special case -- one that suffices for our main interest -- and change notations accordingly. We denote by $S$ and ${\overline S}$ arbitrary sets, and introduce formal variables ${\overline \lambda}_\mybullet:=({\overline \lambda}_s)_{s\in S}$ and  $\lambda_\mybullet:=(\lambda_s)_{s\in {\overline S}}$. The formal Gaussian integral of interest is now 
\[ \int \left(\prod_{i\in J} \frac{d{\overline z}_i\wedge dz_i}{2i\pi\hbar} \exp (- z_i {\overline z}_i/\hbar) \right) \exp (L_{{\overline \lambda}_\mybullet}(z_\mybullet)+ {\overline L}_{{\lambda}_\mybullet}({\overline z}_\mybullet))/\hbar,\]
where $L_{{\overline \lambda}_\mybullet}(z_\mybullet)$ and ${\overline L}_{\lambda_\mybullet}({\overline z}_\mybullet)$ belong to $R[z_\mybullet]({\overline \lambda}_\mybullet)_{\geq 1}$ and $R[{\overline z}_\mybullet](\lambda_\mybullet)_{\geq 1}$ respectively. 
Thus, the ``function'' that we integrate against the Gaussian measure splits as  a product of a ``holomorphic function'' and an ``anti-holomorphic function'', and we are really dealing with a special case of the general discussion.

The semi-classical expansion carries through. The splitting between $z_\mybullet$ and ${\overline z}_\mybullet$ leads to the extremum equations
\[ {\overline V}_i=\frac{\partial L_{{\overline \lambda}_\mybullet}}{\partial z_i} (U_\mybullet) \qquad U_i=\frac{\partial {\overline L}_{\lambda_\mybullet}}{\partial {\overline z}_i} ({\overline V}_\mybullet)\]
with solution $(U_\mybullet^*,{\overline V}_\mybullet^*)$ and extremal value $F^*:= \sum_{i\in J} U_i^*{\overline V}_i^*-L_{{\overline \lambda}_\mybullet}(U_\mybullet^*)-{\overline  L}_{\lambda_\mybullet}({\overline V}_\mybullet^*)$

There is also a (restricted) master integral version adapted to the splitting. The index set $J$ for the variables $(z_\mybullet,{\overline z}_\mybullet)$ is fixed but we consider only monomials purely in $z_\mybullet$ or in ${\overline z}_\mybullet$. We denote now by $S(J)$  the set indexing monomials in $z_\mybullet$  so 
\[ S(J):=\{(m_i)_{i\in J}, \, m_i\in \mathbb{N} \text{ for } i\in J,\, \sum_{i\in J} m_i<+\infty \}.\]
The variables indexed by $S(J)$ are denoted by ${\overline \Lambda}_\mybullet:=({\overline \Lambda}_{m_\mybullet})_{(m_\mybullet)\in S(J)}$ . The set indexing monomials in ${\overline z}_\mybullet$ is denoted by ${\overline S}(J)$ and the variables it indexes by $\Lambda_\mybullet$. We specialize the general graphical rules to compute the restricted master integral 
\[ \int \left(\prod_{i\in J} \frac{d{\overline z}_i\wedge dz_i}{2i\pi\hbar} \exp (- z_i {\overline z}_i/\hbar) \right) \exp (L(J)_{{\overline \Lambda}_\mybullet}(z_\mybullet)+ {\overline L}(J)_{{\Lambda}_\mybullet}({\overline z}_\mybullet))/\hbar,\]
where 
\[ L(J)_{{\overline \Lambda}_\mybullet}(z_\mybullet):=\sum_{m_\mybullet\in S(J)} {\overline \Lambda}_{m_\mybullet}\prod_{i\in J} \frac{z_i^{m_i}}{m_i!} \in R[z_\mybullet]({\overline \Lambda}_\mybullet)_{\geq 1},\]
and 
\[ {\overline L}(J)_{\Lambda_\mybullet}({\overline z}_\mybullet):=\sum_{{\overline m}_\mybullet\in {\overline S}(J)} \Lambda_{{\overline m}_\mybullet} \prod_{i\in J} \frac{{\overline z}_i^{{\overline m}_i}}{{\overline m}_i!}\in R[{\overline z}_\mybullet](\Lambda_\mybullet)_{\geq 1}.\]
Instead of one general family of vertices, we are led to consider two special families of vertices. For $m_\mybullet\in S(J)$ we represent ${\overline \Lambda}_{m_\mybullet} \prod_{i\in J} z_i^{m_i}$ as a \textit{white} vertex with $m_i$ outgoing edges carrying the symbol $i$, while for ${\overline m}_\mybullet\in {\overline S}(J)$ we represent $\Lambda_{{\overline m}_\mybullet} \prod_{i\in J} {\overline z}_i^{{\overline m}_i}$ as a \textit{black} vertex with ${\overline m}_i$ incoming edges carrying the symbol $i$. Thus there is an expansion of the restricted master integral as 
\[ \int \left(\prod_{i\in J} \frac{d{\overline z}_i\wedge dz_i}{2i\pi\hbar} \exp (- z_i {\overline z}_i/\hbar) \right) \exp (L(J)_{{\overline \Lambda}_\mybullet}(z_\mybullet)+ {\overline L}(J)_{{\Lambda}_\mybullet}({\overline z}_\mybullet))/\hbar = \sum_G w(G) \]
where the sum is over all bicolored (white,black) graphs whose edges carry a type $i\in J$. The weight $w(G)$ of a graph is computed as before, though in this restricted case slightly simpler rules could be given to compute the autorphism group. Note that the colors of vertices allow to reconstruct the orientation, edges going from white to black vertices. The restricted master integral exponentiates  as $\exp \sum_G w(G)$ where now the sum is over all connected bicolored graphs whose edges carry a type $i\in J$.

We give a few examples. We take again $J$ to be the standard alphabet $J:=\{a,b,\cdots,y,z\}$. \\
-- Example: A white vertex with $6$ (outgoing) edges, $3$ of type $f$, $2$ of type $y$ and $1$ of type $a$, with weight ${\overline \Lambda}_{a+3f+2y}$:
\begin{center}
\includegraphics[width=.15 \linewidth]{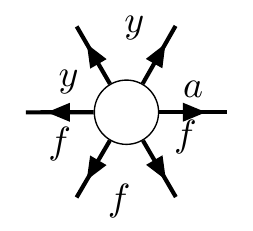}  \raisebox{30pt}{ or simply } \includegraphics[width=.15 \linewidth]{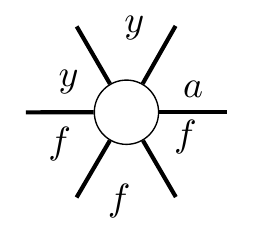} 
\end{center}
-- Example: A black vertex with $5$ (incoming) edges, $2$ of type $r$, $2$ of type $l$ and $1$ of type $u$, with weight $\Lambda_{2{\overline l}+2{\overline r}+{\overline u}}$:
\begin{center}
\includegraphics[width=.15 \linewidth]{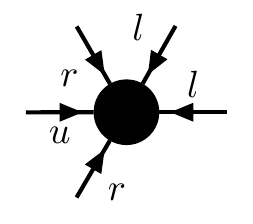}  \raisebox{30pt}{ or simply } \includegraphics[width=.15 \linewidth]{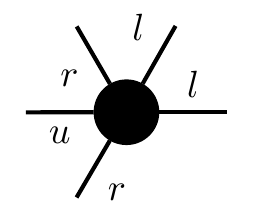} 
\end{center}
-- Example: A diagram
\begin{center}
\includegraphics[width=.4 \linewidth]{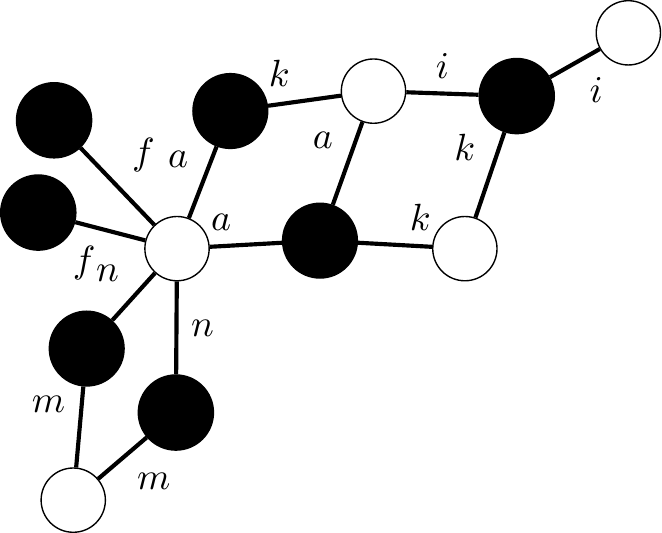} 
\end{center}
with weight (reading more or less from left to right, in that simple case symmetries are ``local'' on the graph)
\[ \frac{1}{2!}\Lambda_{{\overline f}}^2 \frac{1}{2!} {\overline \Lambda}_{2m} \Lambda_{{\overline m}+{\overline n}}^2 {\overline \Lambda}_{2a+2f+2n} \Lambda_{{\overline a}+{\overline k}} \Lambda_{2{\overline a}+{\overline k}} {\overline \Lambda}_{a+i+k} {\overline \Lambda}_{2k} \Lambda_{2{\overline i}+{\overline k}} {\overline \Lambda}_{i}.\]

\end{document}